\documentclass[amsmath,twocolumn,superscriptaddress,prx,aps, longbibliography]{revtex4-1} 
\usepackage{amsthm,amsfonts,graphicx,verbatim, color}
\usepackage{bm}
\usepackage[utf8]{inputenc}
\let\oldmarginpar\marginpar
\renewcommand\marginpar[1]{\-\oldmarginpar[\raggedleft\footnotesize #1]%
{\raggedright\footnotesize #1}}

\newcommand{\be}{\begin{equation}}
\newcommand{\ee}{\end{equation}}
\newcommand{\bea}{\begin{eqnarray}}
\newcommand{\eea}{\end{eqnarray}}

\renewcommand{\epsilon}{\varepsilon}

\newcommand{\jm}[1]{{\color{blue}}}

\def\beq{\begin{equation}}
\def\eeq{\end{equation}}
\def\bea{\begin{eqnarray}}
\def\eea{\end{eqnarray}}

\begin{document}

\title{Many body localization proximity effects in platforms
of coupled spins and bosons}
\author{J. Marino}
\affiliation{Department of Physics and Center for Theory of Quantum Matter, University of Colorado Boulder, Boulder, Colorado 80309, USA}
\author{R.M. Nandkishore}
\affiliation{Department of Physics and Center for Theory of Quantum Matter, University of Colorado Boulder, Boulder, Colorado 80309, USA}

\begin{abstract}
We discuss the onset of  many body localization in a one-dimensional system composed of a  XXZ quantum spin chain  and a Bose-Hubbard model linearly coupled together.
We consider two complementary setups depending whether spatial disorder is initially imprinted on spins or on bosons; in both cases, we explore the conditions for the   disordered portion of the system to localize by proximity of the other  clean half.
%
%
Assuming that the dynamics of one of the two parts develops on shorter time scales than the other, we can adiabatically eliminate the fast degrees of freedom, and derive an effective hamiltonian for the system's remainder using projection operator techniques.
Performing a locator expansion on the strength of the many-body interaction term or on the hopping amplitude of the effective hamiltonian thus derived, we present results on   the stability of the many-body localized phases induced by proximity effect. 
We also briefly comment on the feasibility of the proposed model through modern quantum optics architectures, with the long-term perspective to realize  experimentally, in composite open systems, Anderson or many-body localization proximity effects.

\end{abstract}
\maketitle

\section{Introduction} 
%
%
%
Inhibition of energy transport induced by the presence of spatial disorder has its origins at the middle of last century~\cite{Anderson, AALR}, and its
phenomenology has been confined to free systems till the last decade. 
More recently, a number of works ranging from applications of perturbation theory~\cite{AGKL, BAA, Mirlinrecent} to numerical simulations~\cite{Pal, Znidaric, OganesyanHuse}  have established that a localized phase exhibiting absence of diffusion and equilibration on long time scales, can survive the presence of many body interactions.
For instance, a one dimensional disordered quantum system can show a many-body localized phase~\cite{Nandkishore-2015, VHO, AbaninReview} even at strong interactions or at high energy densities, failing to act as its own bath. 
The existence of a many-body localized  phase has been recently proven rigorously for one-dimensional systems on a lattice   with short-range interactions~\cite{Imbrie}, and explored in experiments with cold gases~\cite{Schreiber2015, Bordia, Choi2016}.
This novel phase of matter not only displays breakdown of ergodicity as salient feature, but it also shows a rich phenomenology, including connections to integrability \cite{Bardarson2012, Serbyn, HNO, Scardicchio, lstarbits, GBN, NonFermiGlasses}, unusual response properties \cite{nonlocal, mblconductivity},  a rich pattern of quantum entanglement \cite{Bardarson, Geraedts2016, KhemaniPRX, Chamon, GRN}, and new types of order that cannot arise in equilibrium  \cite{LPQO, Pekkeretal2014, VoskAltman2014}. This many body localization (MBL) is accordingly drawing considerable interest. 

While much effort has been devoted to the question of when an isolated quantum system can be localized \cite{QHMBL, 2dcontinuum, anycontinuum, proximity, nonabelian, DeRoeck2014, mblmobilityedges, mblbathgeneral, avalanches, SILL, LRMBL}, in reality, any system is unavoidably coupled to an environment, and understanding the interplay of a many-body localized phase communicating with  a thermal bath is of paramount importance, both for the exploration of the phenomenon in  experiments,  as well as a  to understand its robustness to ergodic perturbations.
The naive expectation that the bath can provide sufficient energy and phase-space to facilitate the hopping in an otherwise localized system, has been confirmed by a series of theoretical studies~\cite{NGH, gn, NGADP, BanerjeeAltman, Altmandeph, Levi, Medv16, avalanches, deroeck17, Chandran17}  and by a recent experiment~\cite{Bloch17}.
These works  have however shown  that the interplay of localization and dissipation can leave signatures on the evolution of observables of interest at intermediate times, before thermalization establishes eventually.
%
%

These findings suggest an inherent fragility of disorder-induced localized phases of matter to the coupling to the environment.
However, there exist other captivating scenarios when  open systems supporting a localized phase are not necessarily doomed to a restoration of thermodynamic equilibrium; in addition to dissipation engineering via non-local Lindbladians capable to drive a  system into a  state with desired  localization properties, the possibility of a \emph{many body localization proximity effect} represents a promising direction.  
The phenomenon has been originally  discussed in a toy model of two interacting systems composed of different elementary degrees of freedom~\cite{proximity}:
when a  disordered system is coupled to a clean one with comparable size acting virtually as a bath, the former can in turn, for certain choise of interaction and disorder strengths, induce localization on the latter.
Starting from the analysis in  Ref.~\cite{proximity}, we aim at presenting in this work a realistic setup (see for some experimental proposals Sec.~\ref{concl}) where a many body proximity effect can occur in a quantum spin chain coupled to interacting bosons hopping on a lattice (Sec.~\ref{sec:zero}). Our analysis makes use of the Lindblad formalism, which has not been previously applied to the MBL proximity effect. 
We will consider both  disorder  imprinted on the spin~(Sec.~\ref{sec:uno}) or bosonic sector~(Sec.~\ref{sec:two}), and  establish the conditions for the robustness of a many-body localized phase, induced by the proximity of  the disordered system into the clean one.
We provide a series of estimates for the borders of the localization/delocalization transition in various parameter regimes, basing our study on the   combination of adiabatic elimination techniques and  locator expansion, which can serve as guideline for subsequent numerical and analytical developments.
We also discuss the more natural  occurrence of  the clean, ergodic system acting  as a bath, which delocalizes the disordered one.
In view of proposing viable platforms  to observe these effects, we also suggest at the end of the paper~(Sec.~\ref{concl}) some potentially interesting quantum optics platforms where the phases discussed in this work might be experimentally explored in the future. 

\section{The model} 
\label{sec:zero}
\begin{figure}
 \centering
    \includegraphics[width=8.8cm]{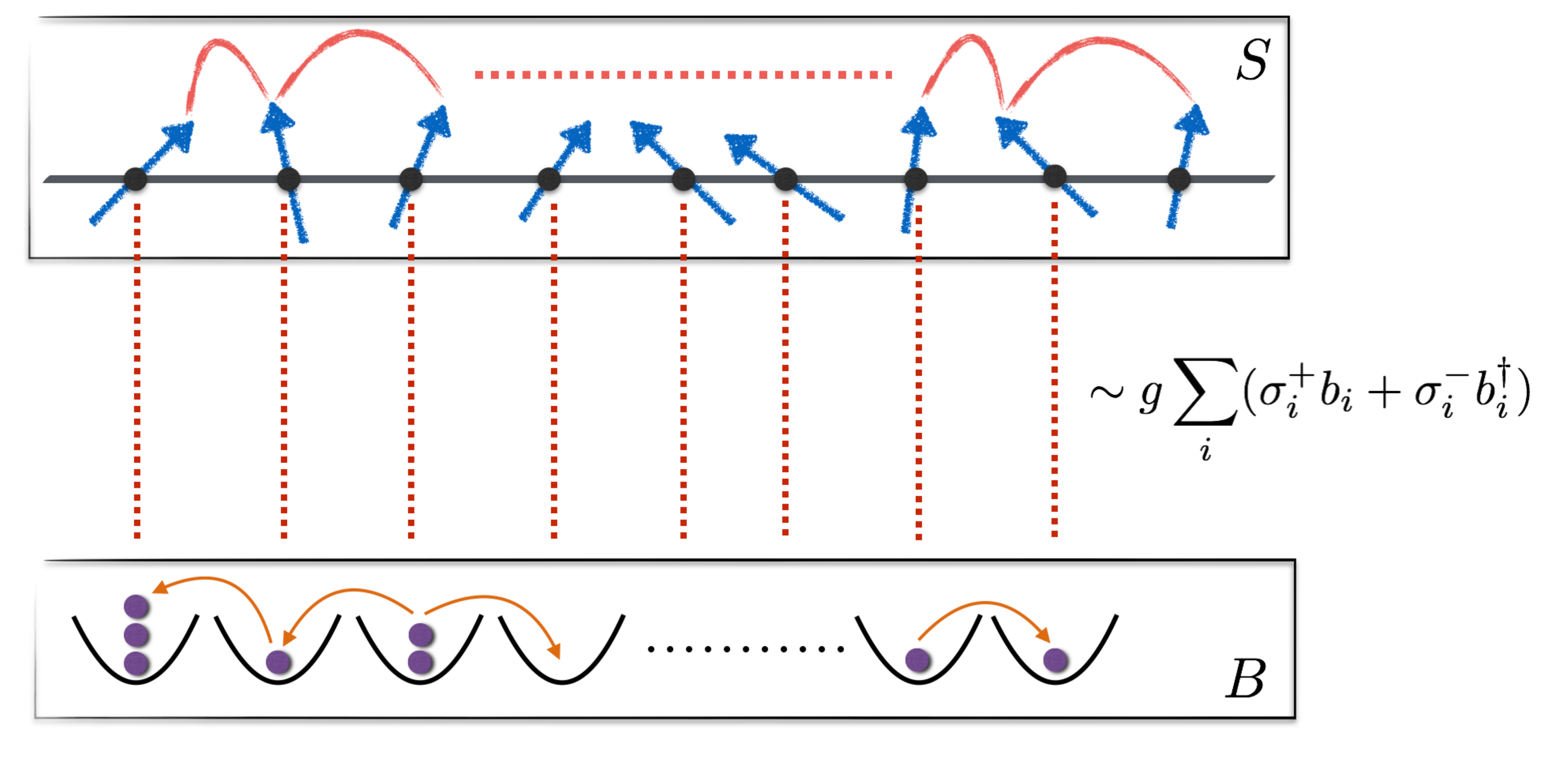}
    \caption{
 We consider an interacting quantum spin chain coupled to a Bose-Hubbard model by a linear, dipole-like interaction term, $H_{int}=g\sum_i (\sigma^+_i~b_i+h.c.)$, see hamiltonian~\eqref{eq:totham}.
When the typical energy scale ruling the dynamics of the bosons (or of the spins) is larger than the coupling $g$, one can adiabatically eliminate the bosonic (spin) sector, and derive an effective hamiltonian for the reminder of the system, \emph{viz.} spins (or bosons). 
We study the  onset of a many body localized phase in these effective hamiltonians derived in adiabatic elimination, imprinting first disorder on the spin sector (S) in Sec.~\ref{sec:uno}, and later in Sec.~\ref{sec:two} on the bosonic (B) one.}
     \label{fig1}
\end{figure}

We consider a one dimensional  model composed of a XXZ quantum spin chain coupled linearly to a Bose-Hubbard model, as given by the hamiltonian (see also Fig.~\ref{fig1})
\begin{equation}
\label{eq:totham}
\begin{split}
H&=-J\sum_{\langle i,j \rangle} b^\dag_i b_j+\frac{U}{2}\sum_i n_i(n_i-1)-\sum_i\omega_ib^\dag_i b_i+\\
&+H_{int} +\alpha\sum_{\langle i,j\rangle}\sigma^+_i\sigma^-_j+\lambda\sum_i\sigma^z_i\sigma^z_{i+1}+\sum_i h_i\sigma^z_i,
\end{split}
\end{equation}
The term $H_{int}$ denotes a spin-boson, dipolar-like (see Sec.~\ref{concl} for the choice of terminology in connection with potential experimental realisations), interconversion term
\begin{equation}
H_{int}=g\sum_i (\sigma^+_i~b_i+h.c.),
\end{equation}
while $n_i=b^\dag_ib_i$ labels bosonic occupation numbers, and $\sigma^\alpha_i$ ($\alpha=x,y,z$) are Pauli matrices describing spins-1/2 on sites $i=1,...,N$;
the sum $\sum_{\langle i,j\rangle} ...$ is intended over next-nearest-neighbours.  
The transverse field $h_i$ and the chemical potential $\omega_i$ are taken site-dependent, since they will alternatively host spatial disorder.

A Jordan-Wigner transformation renders the XXZ quantum spin chain an interacting fermionic model, which, in presence of disorder, constitutes a paradigm for the many body localization transition~\cite{Pal, Znidaric}. 
We therefore first imprint disorder on the transverse fields, $h_i$, drawn accordingly from a uniform probability distribution of width, $W$, and  consider the emergence of a many-body localization transition in  the bosonic chain by proximity of the disordered spin model, Sec.~\ref{sec:uno}. 
Alternatively, we consider on-site disorder (uniformly drawn from a distribution of width $\Omega$) on the chemical potential of bosons, $\omega_i$  (see Sec.~\ref{sec:two}), and study the impact of the latter on the quantum spin chain. 
A schematic representation of the setups considered is provided in Fig.~\ref{fig1}.

In order to estimate  the border of the many-body localized phase, 
we will always consider  a disorder  sufficiently strong to push the disordered sector of the system (spins or bosons) deep into the localized phase  and evaluate its impact on the  latter portion, assumed clean. 
A fully many-body-localized system (i.e. with all the many-body eigenstates localized), renders more transparent the analysis, since the effective hamiltonian can be approximated by a sum of  several single particle terms, with the range of many-body interactions falling exponentially with a short correlation length if disorder is sizeable. Under this hypothesis the conserved, $l-$bits of the many-body localized phase~\cite{HNO, Imbrie, Imbriereview} are close to original bare degrees of freedom employed in hamiltonian~\eqref{eq:totham}, which  can be therefore employed in the subsequent analysis without qualitatively affecting the generality of the conclusions.
%


\section{Disorder in the spin sector} 
\label{sec:uno}
\subsection{MBL proximity effect for bosons}
\label{sec:one}

The first case considered in our analysis  consists of spins subject to  disordered fields, $h_i\in[-W/2,W/2]$, along the $\hat{z}$-direction, and linearly coupled to bosons  as described by the hamiltonian
\begin{equation}\label{eq:ham1}
\begin{split}
H&=-J\sum_{\langle i,j \rangle} b^\dag_i b_j+\frac{U}{2}\sum_i n_i(n_i-1)-\omega_0\sum_ib^\dag_i b_i+\\
&+H_{int} +\sum_i h_i\sigma^z_i.
\end{split}
\end{equation}
Compared to \eqref{eq:totham}, we have assumed a constant chemical potential $\omega_i=\omega_0$, $\forall i=1,...,N$.
The hamiltonian~\eqref{eq:ham1} models the transfer of disorder from a bath of free spins into a clean system of bosons via the linear term $H_{int}$; as we commented in Sec.~\ref{sec:zero},  spins precessing in strong ($W\gg \lambda, \alpha$), disordered transverse fields $h_i$, can  be interpreted as resulting from a  disordered XXZ  spin chain, deep in the  many-body localized phase. In the $W \rightarrow \infty$ limit all other terms in the spin Hamiltonian can be neglected; at large but finite $W$, the spin Hamiltonian can be written in terms of emergent local integrals of motion or `l-bits,' which are approximately equal to the $\sigma^z_i$, with some dressing by higher spin terms ~\cite{HNO, Imbrie, Imbriereview}. 

Within the energy scale separation $g\ll W$, i.e. assuming that the spin bath has  faster dynamics than the bosonic system ($\omega\ll W$), we can integrate out the former and find an effective dynamics for the bosons, using projection operator techniques~\cite{Breuerbook}. 
In particular, denoting with $\rho_b(t)$ the reduced density matrix for the bosons at time $t$, we can write an effective master equation,  where spin dynamics has been averaged out~\cite{Breuerbook, Marcos12}, as a result of  second order perturbation theory in the coupling $g$,
\begin{equation}\label{eq:elimin}
\begin{split}
\dot{\rho}_b(t)&=\mathcal{L}_b\rho_b(t)-\\
&-\int^\infty_0 d\tau \operatorname{Tr}_s\{[H_{int},e^{\tau\mathcal{L}_s}\big([H_{int},\rho_b(t)\otimes\rho^0_s]\big)]\}.
\end{split}
\end{equation}

In Eq.~\eqref{eq:elimin}, the density matrix   $\rho^0_s$ encodes the state in which the spins are  frozen as a result of the above mentioned energy scale separation, and $\mathcal{L}_s$ is the  Liouvillian evolutor of the spin sub-system; finally, $\operatorname{Tr}_s\{...\}$ denotes the trace on the spin's Hilbert space.

Eq.~\eqref{eq:elimin} holds in the generic case of  dynamics with co-existent hamiltonian and Lindbladian processes~\cite{Breuerbook}.
Although for large part of the paper, we will consider purely hamiltonian evolutions, in Sec.~\ref{concl} we study the impact of dissipation on the spin sector, where the use of Eq.~\eqref{eq:elimin} is convenient to perform adiabatic elimination.

In order to apply \eqref{eq:elimin}, we first assume that the bosonic gas is in the atomic limit ($J=0$) and subsequently (see discussion after Eq.~\eqref{eq:hameff1}) we perform a locator expansion in the hopping $J$, in order to find the border of the many body localization transition as a function of the couplings appearing in Eq.~\eqref{eq:ham1}.
The effective hamiltonian, $H'_b$, for the bosonic system is easily identified writing the right hand side of \eqref{eq:elimin} as the generator of a unitary dynamics, 
\begin{equation}
\dot{\rho}_b(t)=-i[H'_b,{\rho}_b(t)].
\end{equation} 
The trace over the spin degrees of freedom is performed assuming that the spin system is in an eigenstate of the disordered spin hamiltonian, $h_s=\sum_ih_i\sigma^z_i$.
The computations on the right hand side of Eq.~\eqref{eq:elimin} can be easily carried (see for instance~\cite{Marcos12, Agarw} and the Appendix), as they require to  freely evolve the  spin operators in $H_{int}$ with $e^{\tau\mathcal{L}_s}$ (in this case, we have simply $\mathcal{L}_s=ih_s$), and to evaluate the time integral in the variable $\tau$. 
Since spin states are localized in real space (as consequence of having assumed the XXZ spin chain deep in the localized phase),  the new effective disordered bosonic term in $H'_b$ will be local in space as well: 
indeed, the matrix element $\sum_{i,j}\langle\sigma^-_i\sigma^+_j\rangle b^\dag_ib_j$, occurring in the evaluation of Eq.~\eqref{eq:elimin}, will have local spatial support, $\langle\sigma^-_i\sigma^+_j\rangle\sim \delta_{i,j}$, because strong disorder rules out   spin  states,  $\rho^0_s$, delocalized in real space.
Eq. \eqref{eq:elimin}, therefore, yields the effective bosonic hamiltonian
\begin{equation}
\label{eq:hameff1}
H'_b= -J\sum_{\langle i,j \rangle} b^\dag_i b_j+\frac{U}{2}\sum_i n_i(n_i-1)-\sum_i\Big(\omega_0+\frac{g^2}{s_ih_i}\Big)n_i,
\end{equation}
where the disordered dressing of the bosonic frequency $\omega_0$ is the main effect  of the adiabatic elimination  of the spins. 
The term $s_i$  in~\eqref{eq:hameff1} is reminiscent of the state of the spin bath, $\rho^0_s$, and it can acquire only the values $s_i=\pm1$, depending whether in $\rho^0_s$ the spin on the lattice site $i$, was respectively oriented up or down along the $\hat{z}$ direction;
however, since $h_i$ is drawn from a distribution with support in $[-W/2,W/2]$ and centred around zero mean, the sign of $s_i$ cannot qualitatively alter the results of the following analysis.
The impact of the hopping term $\propto J$, on the otherwise trivially localized boson hamiltonian (Eq.~\eqref{eq:hameff1} with $J=0$), will now be discussed in locator expansion. 
As a side remark, we notice that  if the  dynamics of  spins is itself dissipative, terms in the Lindblad form will appear in the effective Liouvillian for the bosons as a result of the calculations contained in Eq.~\eqref{eq:elimin}; see for instance  Sec.~\ref{dissip}. 

As anticipated, we now consider the effect of the hopping $J$ in locator expansion. 
A regime of interest is the one where  $\omega_0$ is negligible compared to the disordered correction $\propto g^2$ arising from  adiabatic elimination of spins, $\omega_0\ll{g^2}/{W}$.
At $J=0$, hamiltonian \eqref{eq:hameff1} is trivially localized because in the atomic limit; switching  the hopping, locator expansion~\cite{proximity, QHMBL} predicts a critical value of $J$, at which transport is restored in the system 
\begin{equation}\label{Jcritic}
J\gtrsim J_c\simeq \frac{g^2}{W} \exp(-\frac{1}{2}\Sigma_T\xi).
\end{equation}
In the above expression $\xi$ is the localization length of bosons and $\Sigma_T$ the entropy density at fixed temperature $T$ (we recall that in the estimate~\eqref{Jcritic} and in the others following in the next sections, the localization length $\xi$ is measured in units of the lattice spacing of~\eqref{eq:totham}, and therefore adimensional).
Formula~\eqref{Jcritic} is derived (see Ref.~\cite{proximity}) requiring that the typical matrix element, $J\exp(-\frac{1}{2}\Sigma_T\xi)$, among states connected by the bosonic hopping term is larger than their off-shell energy level spacing, $(g^2/W) \exp(-\Sigma_T\xi)$, indicating the breaking of the locator expansion, and the onset of a delocalized ergodic phase in the system. 
However, as long as  $J\lesssim J_c$, the bosonic system described by \eqref{eq:hameff1} will be in a localized phase induced by the coupling to the original stronlgy disordered spin model and representing accordingly a typical  instance of \emph{many-body localization proximity effect}.

\subsection{Delocalization of the disordered spin bath}
\label{sec:dueA}

We now consider a regime of  energy scales separation complementary to the one discussed above. 
Specifically, we set to zero the spin hopping term ($\alpha=0$ in Eq.~\eqref{eq:totham}), and discuss the  generation of a next-neighbour spin-flipping term when the clean bosonic system is traced out.
The impact of the transverse spin-spin interaction term $\propto\lambda$ will be analysed in locator expansion.
When $g\ll\omega_0$ in  hamiltonian~\eqref{eq:ham1}, we can  adiabatically eliminate the bosons, if they evolve faster compared to spins, $\omega_0\gg W$, and derive an effective hamiltonian for the latter using once again the formula \eqref{eq:elimin} where the role of bosons and spins are now interchanged with respect to Sec.~\ref{sec:one}.
Since the bosons are  clean from disorder, we can now assume that they are frozen in a delocalized state in real space, contrary to the '\emph{l}-bits' of Sec.~\ref{sec:one}; this carries the consequence that the effective spin hamiltonian can inherit an hopping term $\propto g^2\sum_{\langle i,j \rangle}\sigma^+_i\sigma^-_{j}$ after bosons are traced away, because the matrix element $\langle b_i b^\dag_j \rangle$ can have non-local support on the lattice.

The effective hamiltonian for the spins degrees of freedom reads then,
\begin{equation}\label{eq:hopspin}
H'_s=\sum h_i\sigma^z_i+\sum_{ i}\frac{g^2}{\omega_0+U(1+2\bar{n}_i)}(\sigma^+_i\sigma^-_{i+1}+h.c.).
\end{equation}
For instance, in order to generate a hopping to next neighbouring sites as in Eq.~\eqref{eq:hopspin}, is sufficient to trace out a bosonic state, $|\psi\rangle_b$, written as a superposition $|\psi\rangle_b=c_1|\{01\}_i\rangle+c_2|\{10\}_i\rangle$ (with $c_1$ and $c_2$  arbitrary constants), where the shorthand $|\{01\}_i\rangle$  stands for a state with sites alternatively occupied and unoccupied by bosons, $|010101...01\rangle$, and the state $|\{10\}_i\rangle=|10101...10\rangle$ is just obtained by a unit shift of the lattice site position in $|\{01\}_i\rangle$. 
The number $\bar{n}_i$ in \eqref{eq:hopspin} is the expectation value of the conserved, local bosonic number $\hat{n}_i$ (we recall that we considering the atomic limit $J=0$ for bosons).

A one dimensional, non-interacting disordered system is an Anderson insulator at any disorder strength; in order to study the potential onset of delocalization in this system via many body spin-spin interactions, we treat in locator expansion the  term
\begin{equation}
V=\lambda \sum_i \sigma^z_i\sigma^z_{i+1}
\end{equation} 
in the hamiltonian~\eqref{eq:hopspin} (therefore now $H'_s$ has been dressed by $V$), having in mind the structure of  the XXZ spin chain  in \eqref{eq:totham}.
Since $V$ is a quartic fermionic interaction  after Jordan-Wigner fermionization, we can estimate the magnitude of the typical term of the   locator series~\cite{QHMBL,proximity} connecting two localized states $|\alpha\rangle$ and $|\beta\rangle$ of the spin hamiltonian~\eqref{eq:hopspin} with energies respectively $E_\alpha$ and $E_\beta$,
\begin{equation}\label{eq:ratioloc}
\frac{\langle\alpha| V |\beta\rangle}{E_\alpha-E_\beta}\sim\frac{\lambda \xi^{-d}}{J'\xi^{-4d}}
\end{equation}
where $d=1$ in our case and $J'\sim g^2/(\omega_0+U)$ is the effectively generated hopping in hamiltonian~\eqref{eq:hopspin}.
The estimate for the matrix element of the interaction $\langle\alpha| V |\beta\rangle$ is derived~\cite{QHMBL} in the basis of the localized wave functions of the spins, and requires to take properly into account  the  normalization of the localized states, $\propto\xi^{-d/2}$, while the energy level spacing in the denominator, $\Delta E=E_\alpha-E_\beta$, is proportional to $\xi^{-4d}$, since the interaction involves four localized fermions, acting each one on a region of order $\sim \xi^{-d}$. 
Since the model is one dimensional, locator expansion for the free model~\eqref{eq:hopspin} predicts a localization length smaller than lattice spacing and scaling as the logarithm of the ratio between disorder and hopping strengths, $\xi\propto 1/\log\left(W/J'\right)$, since the spin chain~\eqref{eq:hopspin} will be in a strongly localized regime for the energy scale separation adopted.
Therefore, the  threshold in the coupling $\lambda$ for delocalizing the disordered spin bath is, using Eq.~\eqref{eq:ratioloc}, 
\begin{equation}
\lambda\gtrsim \lambda_c\simeq J'\log^3(W/J'),
\end{equation} 
since as far as $\lambda\lesssim \lambda_c$ and the locator expansion  converges, the eigenstates of the interacting problem would remain close to those of the Anderson insulator~\eqref{eq:hopspin} present at $\lambda=0$.
%
%

In a similar fashion, if we  start with  bosons in a hopping dominated regime ($U=0$, $t\neq0$) and with spins in a strongly localized phase, the adiabatic elimination of the latter will imprint on-site disorder on the former which will immediately localize the bosons at any disorder strength (once again, the system is one dimensional). 
If one treats on-site bosonic interactions in locator expansion, one can easily show, following the same procedure outlined above, that the spin-induced, localized phase of bosons melts as  interaction's strength, $U$, reaches values  $\propto t\log^3\left({g^2}/({Wt})\right)$, where we have employed again  strong localization theory.


\section{Disorder in the bosonic sector}
\label{sec:two}
\subsection{MBL proximity effect for spins}
The setup of the second part of this work complements the one of the previous Section. 
Disorder is now imprinted on bosons, specifically, the chemical potentials $\omega_i$ are drawn from a uniform distribution of width $\Omega$ and with zero mean value; complementing Sec.~\ref{sec:one}, we aim at inspecting the impact of the disorder transfer  from bosons to the quantum spin chain hamiltonian with uniform transverse fields, $h_i=h_0$, cfr.~Eq.~\eqref{eq:totham},
\begin{equation}
\begin{split}
H=&\sum_i h_0\sigma^z_i+\lambda\sum_i\sigma^z_i\sigma^z_{i+1}+\alpha\sum_{\langle i,j\rangle}\sigma^+_i\sigma^-_j +\\ 
&+H_{int} +\sum_i\omega_ib^\dag_ib_i.
\end{split}
\end{equation} 
Since we now wish to adiabatically eliminate the  bosons, we adopt  the energy scales separation $g\ll\Omega$ and $h_0\ll\Omega$, while
hopping will be subsequently switched on in locator expansion.  
The choice of a diagonal bosonic hamiltonian can, once again, be considered the extreme limit of a strongly disordered, interacting bosonic system, deep in its many-body localized phase (see the discussion in Sec.~\ref{sec:zero} and Refs.~\cite{HNO, Imbrie, Imbriereview}).
In the spin 'atomic' limit $\alpha\to0$ ($\sigma^z_i$ is conserved at each lattice site), the calculations necessary to adiabatically eliminate bosons using Eq.~\eqref{eq:elimin}, can be straightforwardly performed.
Assuming the bosons frozen into a localized state, we find  that they contribute with a disordered shift to the constant transverse field $h_0$, in the effective spin hamiltonian
\begin{equation}\label{eq:spin4}
H'_s=\sum_i \Big(h_0+\frac{g^2}{\omega_i}\Big)\sigma^z_i+\lambda\sum_i\sigma^z_i\sigma^z_{i+1}+\alpha\sum_{\langle i,j\rangle}\sigma^+_i\sigma^-_j.
\end{equation}
Considering in this case the hopping as a perturbation $V=\alpha\sum_{\langle i,j\rangle}\sigma^+_i\sigma^-_j$, capable to connect states which otherwise would be  localized, we can estimate the magnitude of the characteristic term of the locator expansion
\begin{equation}
\frac{\langle\alpha| V |\beta\rangle}{E_\alpha-E_\beta}\sim\frac{\alpha}{g^2/\Omega \exp(-s_T\xi)},
\end{equation}
using an argument analogous to the one employed for~\eqref{Jcritic}.
We can then conclude that the series would converge  for 
\begin{equation}
\alpha \lesssim g^2/\Omega \exp(-\Sigma_T\xi),
\end{equation}
 and the system's eigenstates would remain close to the localized ones (as in Eq.~\eqref{Jcritic}, $\Sigma_T$ is the entropy density at temperature $T$).
It is straightforward to notice the analogies between the hamiltonians ~\eqref{eq:hameff1} and ~\eqref{eq:spin4}, and their respective results.

\subsection{Delocalization of the disordered bosonic bath}

In the complementary limit of  spin dynamics occurring on  faster time scales than the bosonic one, $g\ll h_0$, $\Omega \ll h_0 $, we find that the impact of the clean, spin bath is to restore  ergodicity in the disordered boson hamiltonian.
As in Sec.~\ref{sec:dueA}, we assume that there is no bare bosonic hopping in \eqref{eq:totham}, and integrate out  the spin sector frozen in a delocalized state, deriving an effective tunnelling  for bosons.
Indeed, the effective bosonic hamiltonian reads in this case
\begin{equation}
H'_b=\sum_i\omega_in_i-\sum_{\langle i,j\rangle}\frac{g^2}{h_0}(b^\dag_ib_j+h.c),
\end{equation}
which again is a model of disordered free particles in one dimension,  localized at any disorder strength $\Omega$. 
A locator expansion using the on-site many-body bosonic interaction term as perturbation
\begin{equation}
V=U\sum_in_i(n_i-1),
\end{equation}
 leads to the estimate
 \begin{equation}\label{eq:critic}
 U_c\sim \frac{g^2}{h_0}\log^3\Big(\frac{\Omega h_0}{g^2}\Big),
 \end{equation}
for the magnitude of the critical interaction strength for delocalization, where in Eq.~\eqref{eq:critic} we have employed the expression of $\xi$ for a strongly disordered system one dimension, $\xi\propto \log\Big(\frac{\Omega h_0}{g^2}\Big)$, replacing the effective hopping $\propto g^2/h_0$ (compare with Sec.~\ref{sec:dueA}).

Analogue results to those discussed at the end of Sec.~\ref{sec:dueA}, are found if we consider the bosons in the localized phase coupled to  a XX-quantum spin chain ($\lambda=0$), and treating the spin-spin interaction along the $\hat{z}$ direction in locator expansion. 
As expected, once again, the latter gets in turn localized by the former: eliminating the bosons imprints on-site disorder on the spins, which gets localized until the interaction strength $\lambda$ ramps up to a threshold which would restore ergodicity in the system. 

\subsection{Coupling to thermal spins and dissipation}
\label{dissip}
The generality of Eq.~\eqref{eq:elimin} allows us to adiabatically eliminate systems with both coherent and dissipative dynamics.
This offers the opportunity to briefly comment on the fate of the Bose Hubbard model in~\eqref{eq:ham1} with on-site disorder of strength $\propto\Omega$ (as the one considered through this section), in contact with a system of spins at equilibrium with a thermal bath at temperature $T$.
This can be, for instance, achieved adding to the simple spin hamiltonian in~\eqref{eq:ham1}, on-site  Lindblad operators
\begin{equation}
L_i=\sigma^+_i,\quad L_i=\sigma^-_i,
\end{equation}
respectively with rates $\gamma_+=\gamma n_T$ and $\gamma_-=\gamma (n_T+1)$, where $n_T$ is the spin thermal distribution at  frequency $h_0$ and  temperature $T$.
We assume the scale separation $h_0$, $\gamma_{\pm}\gg g$, and  consider the spins  frozen in their thermal state at temperature $T$.
The adiabatic elimination of the spin sector adds an effective dissipation to the bosons (see for instance~\cite{Agarw}), described by the dephasing jump operator, $\mathcal{L}_i=b^\dag_ib_i$, and occurring  at rate $\Gamma=\tilde{\gamma} g^2/(\tilde{\gamma}^2+h^2_0)$ where $\tilde{\gamma}=\gamma/\tanh(h_0/2T)$; the   contribution to the bosonic hamiltonian is inconsequential, since the thermal state $\rho_T\sim e^{- \sum_ih_0\sigma^z_i/T}$ is factorised into a product of local density matrices at each site, yielding a constant energy level shift to the disordered chemical potential, $\omega_i$.

Unless  $h_0\gg\tilde{\gamma}$,  which would render the dissipation rate $\Gamma$ small, we arrive at the natural conclusion that, on  time scales of the order $\sim1/\Gamma$, the localized phase melts under the effect of the dephasing operators $\mathcal{L}_i$, without need to invoke a delocalization transition mediated by coherent many body interactions.

\section{Perspectives}  
\label{concl}
\subsection{Experimental implementations}
\label{exps}
The model~\eqref{eq:totham} discussed in this work originates from the coupling of two paradigmatic hamiltonians of condensed matter physics. 
In this Section, we briefly outline possible realizations of the many-body localization proximity effects presented here, from the perspective  of quantum many body simulation using quantum optics platforms.

A first simple experimental  implementation of our set-up consists in considering  circuit QED-arrays~\cite{houck12}, modelling a Jaynes-Cummings lattice with nearest neighbour photon hopping.
The latter consists of a one-dimensional array of  micro-wave resonators supporting photonic excitations at a given frequency  set by the cavity modes;
a linear dipole interaction of the type discussed in this work, $H_{int}$, converts bosons into excitations of superconducting qubits (spins one-half), each of them coupled  to one of the resonators.
Tunnelling among nearby photonic cavities implements a kinetic hopping term of the Hubbard type, leading to a competition, even in the clean system, among  delocalization of photons, and the trapping of the associated polaritons, arising from effective on-site interactions~\cite{houck12}.
In turn, the phase diagrams of the Jaynes-Cummings and of the Bose-Hubbard model bear several similarities~\cite{houck12}, and it would be of interest to explore, in the laboratory, the impact of disorder present in individual cavity frequencies~\cite{Underwood12}, using as guidelines the prediction of this work.
Indeed, the increasing scalability of circuit QED lattice experiments with the consequent possibility to observe phase transitions~\cite{Kirk17}, make in principle these hybrid qubit-photon architectures promising grounds to realise  many-body localization proximity effects.

A more complex, yet closer,  platform to implement the ideas presented in this work, are bosonic Hubbard models of photons on a lattice~\cite{Hartmann, Tomadin, Chang} (equipped with effective two-body photon-photon interactions arising from Kerr non-linearities),  coupled to superconducting quibts with a photon-spin dipole interaction term; varying the photonic (or spin) modes from cavity to cavity (or qubit), one can  simulate disorder in this system.
Furthermore, when qubits are placed next to each other, electrostatic or magnetic interactions among them become relevant, mimicking the spin-spin interaction terms in  our starting hamiltonian~\eqref{eq:totham}. 

Another possibility to start with intrinsic spin interactions, as a term of the type $\sim\sum_i\sigma^z_i\sigma^z_{i+1}$, are Rydberg-dressed spin lattices coupled to interacting photons~\cite{spinrdyd}. However, off-diagonal spin couplings terms,  $\propto\sum_i\sigma^+_i\sigma^-_{i+1}$, would be effectively generated only at higher orders in perturbation theory, changing some of the estimates contained in the previous Sections.\\

\subsection{Future directions} 
Our results concerning the onset of many-body localization proximity effects in physically feasible  platforms, are based on a combination of locator expansion and adiabatic elimination methods, and constitute a  promising first route towards a systematic study of many-body localization in realistic systems made up of different species. 
As a first direction, we foresee a study of Anderson localization in a disordered Jaynes-Cummings lattice (see also discussion at the beginning of Sec.~\ref{exps}) using methods suited for one dimensional disordered systems, such as strong disorder renormalization group.
At the next level of complexity, there is the challenging possibility to derive  the effective strong disorder hamiltonian for the many-body spin-boson model in~\eqref{eq:totham} using flow equation methods~\cite{Pekker17}, which have already proven successful~\cite{Thomson17} in the  description of the paradigmatic many-body localization transition of the XXZ  quantum spin chain.
Another promising avenue  consists in going beyond our perturbative estimates with a real-space renormalisation group procedure~\cite{das, fis}; for instance Ref.~\cite{vosk} adopts, as in our work,  an energy scale separation to derive as well an effective hamiltonian, and extract from the latter a flow of the probability distribution of the random couplings of a disordered, interacting quantum spin chain.
This approach might lead to clearer characterisation of the many-body localized phases induced by proximity effect, which up to now has been characterised only through locator expansion techniques.

Finally, an agenda for future studies of the many-body localization transition in quantum optics, would require a systematic inclusion of dissipation since  experimental implementations would be plagued at long times by photon losses or spin decoherence/dephasing terms (see for instance Sec.~\ref{dissip}), which would eventually  spoil the features of the localized phase.\\

\emph{Acknowledgements--} We thank I. Carusotto, R. Lewis-Swan and A. M. Rey for discussions on related topics.
This material is based upon work supported by the Air Force Office of Scientific Research under award
number FA9550-17-1-0183.

\bibliography{LRMBL}

\begin{thebibliography}{69}%
\makeatletter
\providecommand \@ifxundefined [1]{%
 \@ifx{#1\undefined}
}%
\providecommand \@ifnum [1]{%
 \ifnum #1\expandafter \@firstoftwo
 \else \expandafter \@secondoftwo
 \fi
}%
\providecommand \@ifx [1]{%
 \ifx #1\expandafter \@firstoftwo
 \else \expandafter \@secondoftwo
 \fi
}%
\providecommand \natexlab [1]{#1}%
\providecommand \enquote  [1]{``#1''}%
\providecommand \bibnamefont  [1]{#1}%
\providecommand \bibfnamefont [1]{#1}%
\providecommand \citenamefont [1]{#1}%
\providecommand \href@noop [0]{\@secondoftwo}%
\providecommand \href [0]{\begingroup \@sanitize@url \@href}%
\providecommand \@href[1]{\@@startlink{#1}\@@href}%
\providecommand \@@href[1]{\endgroup#1\@@endlink}%
\providecommand \@sanitize@url [0]{\catcode `\\12\catcode `\$12\catcode
  `\&12\catcode `\#12\catcode `\^12\catcode `\_12\catcode `\%12\relax}%
\providecommand \@@startlink[1]{}%
\providecommand \@@endlink[0]{}%
\providecommand \url  [0]{\begingroup\@sanitize@url \@url }%
\providecommand \@url [1]{\endgroup\@href {#1}{\urlprefix }}%
\providecommand \urlprefix  [0]{URL }%
\providecommand \Eprint [0]{\href }%
\providecommand \doibase [0]{http://dx.doi.org/}%
\providecommand \selectlanguage [0]{\@gobble}%
\providecommand \bibinfo  [0]{\@secondoftwo}%
\providecommand \bibfield  [0]{\@secondoftwo}%
\providecommand \translation [1]{[#1]}%
\providecommand \BibitemOpen [0]{}%
\providecommand \bibitemStop [0]{}%
\providecommand \bibitemNoStop [0]{.\EOS\space}%
\providecommand \EOS [0]{\spacefactor3000\relax}%
\providecommand \BibitemShut  [1]{\csname bibitem#1\endcsname}%
\let\auto@bib@innerbib\@empty
\bibitem [{\citenamefont {Anderson}(1958)}]{Anderson}%
  \BibitemOpen
  \bibfield  {author} {\bibinfo {author} {\bibfnamefont {P.~W.}\ \bibnamefont
  {Anderson}},\ }\bibfield  {title} {\enquote {\bibinfo {title} {Absence of
  diffusion in certain random lattices},}\ }\href {\doibase
  10.1103/PhysRev.109.1492} {\bibfield  {journal} {\bibinfo  {journal} {Phys.
  Rev.}\ }\textbf {\bibinfo {volume} {109}},\ \bibinfo {pages} {1492--1505}
  (\bibinfo {year} {1958})}\BibitemShut {NoStop}%
\bibitem [{\citenamefont {Abrahams}\ \emph {et~al.}(1979)\citenamefont
  {Abrahams}, \citenamefont {Anderson}, \citenamefont {Licciardello},\ and\
  \citenamefont {Ramakrishnan}}]{AALR}%
  \BibitemOpen
  \bibfield  {author} {\bibinfo {author} {\bibfnamefont {E.}~\bibnamefont
  {Abrahams}}, \bibinfo {author} {\bibfnamefont {P.~W.}\ \bibnamefont
  {Anderson}}, \bibinfo {author} {\bibfnamefont {D.~C.}\ \bibnamefont
  {Licciardello}}, \ and\ \bibinfo {author} {\bibfnamefont {T.~V.}\
  \bibnamefont {Ramakrishnan}},\ }\bibfield  {title} {\enquote {\bibinfo
  {title} {Scaling theory of localization: Absence of quantum diffusion in two
  dimensions},}\ }\href {\doibase 10.1103/PhysRevLett.42.673} {\bibfield
  {journal} {\bibinfo  {journal} {Phys. Rev. Lett.}\ }\textbf {\bibinfo
  {volume} {42}},\ \bibinfo {pages} {673--676} (\bibinfo {year}
  {1979})}\BibitemShut {NoStop}%
\bibitem [{\citenamefont {Altshuler}\ \emph {et~al.}(1997)\citenamefont
  {Altshuler}, \citenamefont {Gefen}, \citenamefont {Kamenev},\ and\
  \citenamefont {Levitov}}]{AGKL}%
  \BibitemOpen
  \bibfield  {author} {\bibinfo {author} {\bibfnamefont {Boris~L.}\
  \bibnamefont {Altshuler}}, \bibinfo {author} {\bibfnamefont {Yuval}\
  \bibnamefont {Gefen}}, \bibinfo {author} {\bibfnamefont {Alex}\ \bibnamefont
  {Kamenev}}, \ and\ \bibinfo {author} {\bibfnamefont {Leonid~S.}\ \bibnamefont
  {Levitov}},\ }\bibfield  {title} {\enquote {\bibinfo {title} {Quasiparticle
  lifetime in a finite system: A non-perturbative approach},}\ }\href {\doibase
  10.1103/PhysRevLett.78.2803} {\bibfield  {journal} {\bibinfo  {journal}
  {Phys. Rev. Lett.}\ }\textbf {\bibinfo {volume} {78}},\ \bibinfo {pages}
  {2803--2806} (\bibinfo {year} {1997})}\BibitemShut {NoStop}%
\bibitem [{\citenamefont {Basko}\ \emph {et~al.}(2006)\citenamefont {Basko},
  \citenamefont {Aleiner},\ and\ \citenamefont {Altshuler}}]{BAA}%
  \BibitemOpen
  \bibfield  {author} {\bibinfo {author} {\bibfnamefont {D.M.}\ \bibnamefont
  {Basko}}, \bibinfo {author} {\bibfnamefont {I.L.}\ \bibnamefont {Aleiner}}, \
  and\ \bibinfo {author} {\bibfnamefont {B.L.}\ \bibnamefont {Altshuler}},\
  }\bibfield  {title} {\enquote {\bibinfo {title} {Metal--insulator transition
  in a weakly interacting many-electron system with localized single-particle
  states},}\ }\href {\doibase 10.1016/j.aop.2005.11.014} {\bibfield  {journal}
  {\bibinfo  {journal} {Annals of Physics}\ }\textbf {\bibinfo {volume}
  {321}},\ \bibinfo {pages} {1126 -- 1205} (\bibinfo {year}
  {2006})}\BibitemShut {NoStop}%
\bibitem [{\citenamefont {Gornyi}\ \emph {et~al.}(2007)\citenamefont {Gornyi},
  \citenamefont {Mirlin},\ and\ \citenamefont {Polyakov}}]{Mirlinrecent}%
  \BibitemOpen
  \bibfield  {author} {\bibinfo {author} {\bibfnamefont {I.~V.}\ \bibnamefont
  {Gornyi}}, \bibinfo {author} {\bibfnamefont {A.~D.}\ \bibnamefont {Mirlin}},
  \ and\ \bibinfo {author} {\bibfnamefont {D.~G.}\ \bibnamefont {Polyakov}},\
  }\bibfield  {title} {\enquote {\bibinfo {title} {Electron transport in a
  disordered luttinger liquid},}\ }\href {\doibase 10.1103/PhysRevB.75.085421}
  {\bibfield  {journal} {\bibinfo  {journal} {Phys. Rev. B}\ }\textbf {\bibinfo
  {volume} {75}},\ \bibinfo {pages} {085421} (\bibinfo {year}
  {2007})}\BibitemShut {NoStop}%
\bibitem [{\citenamefont {Pal}\ and\ \citenamefont {Huse}(2010)}]{Pal}%
  \BibitemOpen
  \bibfield  {author} {\bibinfo {author} {\bibfnamefont {Arijeet}\ \bibnamefont
  {Pal}}\ and\ \bibinfo {author} {\bibfnamefont {David~A.}\ \bibnamefont
  {Huse}},\ }\bibfield  {title} {\enquote {\bibinfo {title} {Many-body
  localization phase transition},}\ }\href {\doibase
  10.1103/PhysRevB.82.174411} {\bibfield  {journal} {\bibinfo  {journal} {Phys.
  Rev. B}\ }\textbf {\bibinfo {volume} {82}},\ \bibinfo {pages} {174411}
  (\bibinfo {year} {2010})}\BibitemShut {NoStop}%
\bibitem [{\citenamefont {Znidaric}\ \emph {et~al.}(2008)\citenamefont
  {Znidaric}, \citenamefont {Prosen},\ and\ \citenamefont
  {Prelovsek}}]{Znidaric}%
  \BibitemOpen
  \bibfield  {author} {\bibinfo {author} {\bibfnamefont {M.}~\bibnamefont
  {Znidaric}}, \bibinfo {author} {\bibfnamefont {T.}~\bibnamefont {Prosen}}, \
  and\ \bibinfo {author} {\bibfnamefont {P.}~\bibnamefont {Prelovsek}},\
  }\bibfield  {title} {\enquote {\bibinfo {title} {Many-body localization in
  the heisenberg xxz magnet in a random field},}\ }\href {\doibase
  10.1103/PhysRevB.77.064426} {\bibfield  {journal} {\bibinfo  {journal} {Phys.
  Rev. B}\ }\textbf {\bibinfo {volume} {77}},\ \bibinfo {pages} {064426}
  (\bibinfo {year} {2008})}\BibitemShut {NoStop}%
\bibitem [{\citenamefont {Oganesyan}\ and\ \citenamefont
  {Huse}(2007)}]{OganesyanHuse}%
  \BibitemOpen
  \bibfield  {author} {\bibinfo {author} {\bibfnamefont {Vadim}\ \bibnamefont
  {Oganesyan}}\ and\ \bibinfo {author} {\bibfnamefont {David~A.}\ \bibnamefont
  {Huse}},\ }\bibfield  {title} {\enquote {\bibinfo {title} {Localization of
  interacting fermions at high temperature},}\ }\href {\doibase
  10.1103/PhysRevB.75.155111} {\bibfield  {journal} {\bibinfo  {journal} {Phys.
  Rev. B}\ }\textbf {\bibinfo {volume} {75}},\ \bibinfo {pages} {155111}
  (\bibinfo {year} {2007})}\BibitemShut {NoStop}%
\bibitem [{\citenamefont {Nandkishore}\ and\ \citenamefont
  {Huse}(2015)}]{Nandkishore-2015}%
  \BibitemOpen
  \bibfield  {author} {\bibinfo {author} {\bibfnamefont {Rahul}\ \bibnamefont
  {Nandkishore}}\ and\ \bibinfo {author} {\bibfnamefont {David~A.}\
  \bibnamefont {Huse}},\ }\bibfield  {title} {\enquote {\bibinfo {title}
  {Many-body localization and thermalization in quantum statistical
  mechanics},}\ }\href {\doibase 10.1146/annurev-conmatphys-031214-014726}
  {\bibfield  {journal} {\bibinfo  {journal} {Annual Review of Condensed Matter
  Physics}\ }\textbf {\bibinfo {volume} {6}},\ \bibinfo {pages} {15--38}
  (\bibinfo {year} {2015})}\BibitemShut {NoStop}%
\bibitem [{\citenamefont {Vosk}\ \emph {et~al.}(2015)\citenamefont {Vosk},
  \citenamefont {Huse},\ and\ \citenamefont {Altman}}]{VHO}%
  \BibitemOpen
  \bibfield  {author} {\bibinfo {author} {\bibfnamefont {Ronen}\ \bibnamefont
  {Vosk}}, \bibinfo {author} {\bibfnamefont {David~A.}\ \bibnamefont {Huse}}, \
  and\ \bibinfo {author} {\bibfnamefont {Ehud}\ \bibnamefont {Altman}},\
  }\bibfield  {title} {\enquote {\bibinfo {title} {Theory of the many-body
  localization transition in one-dimensional systems},}\ }\href {\doibase
  10.1103/PhysRevX.5.031032} {\bibfield  {journal} {\bibinfo  {journal} {Phys.
  Rev. X}\ }\textbf {\bibinfo {volume} {5}},\ \bibinfo {pages} {031032}
  (\bibinfo {year} {2015})}\BibitemShut {NoStop}%
\bibitem [{\citenamefont {{Abanin}}\ and\ \citenamefont
  {{Papi{\'c}}}(2017)}]{AbaninReview}%
  \BibitemOpen
  \bibfield  {author} {\bibinfo {author} {\bibfnamefont {D.~A.}\ \bibnamefont
  {{Abanin}}}\ and\ \bibinfo {author} {\bibfnamefont {Z.}~\bibnamefont
  {{Papi{\'c}}}},\ }\bibfield  {title} {\enquote {\bibinfo {title} {{Recent
  progress in many-body localization}},}\ }\href@noop {} {\bibfield  {journal}
  {\bibinfo  {journal} {ArXiv e-prints}\ } (\bibinfo {year} {2017})},\ \Eprint
  {http://arxiv.org/abs/1705.09103} {arXiv:1705.09103 [cond-mat.dis-nn]}
  \BibitemShut {NoStop}%
\bibitem [{\citenamefont {Imbrie}(2016)}]{Imbrie}%
  \BibitemOpen
  \bibfield  {author} {\bibinfo {author} {\bibfnamefont {John~Z.}\ \bibnamefont
  {Imbrie}},\ }\bibfield  {title} {\enquote {\bibinfo {title} {Diagonalization
  and many-body localization for a disordered quantum spin chain},}\ }\href
  {\doibase 10.1103/PhysRevLett.117.027201} {\bibfield  {journal} {\bibinfo
  {journal} {Phys. Rev. Lett.}\ }\textbf {\bibinfo {volume} {117}},\ \bibinfo
  {pages} {027201} (\bibinfo {year} {2016})}\BibitemShut {NoStop}%
\bibitem [{\citenamefont {Schreiber}\ \emph {et~al.}(2015)\citenamefont
  {Schreiber}, \citenamefont {Hodgman}, \citenamefont {Bordia}, \citenamefont
  {Luschen}, \citenamefont {Fischer}, \citenamefont {Vosk}, \citenamefont
  {Altman}, \citenamefont {Schneider},\ and\ \citenamefont
  {Bloch}}]{Schreiber2015}%
  \BibitemOpen
  \bibfield  {author} {\bibinfo {author} {\bibfnamefont {M.}~\bibnamefont
  {Schreiber}}, \bibinfo {author} {\bibfnamefont {S.~S.}\ \bibnamefont
  {Hodgman}}, \bibinfo {author} {\bibfnamefont {P.}~\bibnamefont {Bordia}},
  \bibinfo {author} {\bibfnamefont {H.~P.}\ \bibnamefont {Luschen}}, \bibinfo
  {author} {\bibfnamefont {M.~H.}\ \bibnamefont {Fischer}}, \bibinfo {author}
  {\bibfnamefont {R.}~\bibnamefont {Vosk}}, \bibinfo {author} {\bibfnamefont
  {E.}~\bibnamefont {Altman}}, \bibinfo {author} {\bibfnamefont
  {U.}~\bibnamefont {Schneider}}, \ and\ \bibinfo {author} {\bibfnamefont
  {I.}~\bibnamefont {Bloch}},\ }\bibfield  {title} {\enquote {\bibinfo {title}
  {Observation of many-body localization of interacting fermions in a
  quasi-random optical lattice},}\ }\href {\doibase 10.1126/science.aaa7432}
  {\bibfield  {journal} {\bibinfo  {journal} {Science}\ }\textbf {\bibinfo
  {volume} {349}},\ \bibinfo {pages} {842--845} (\bibinfo {year}
  {2015})}\BibitemShut {NoStop}%
\bibitem [{\citenamefont {Bordia}\ \emph {et~al.}(2016)\citenamefont {Bordia},
  \citenamefont {L\"uschen}, \citenamefont {Hodgman}, \citenamefont
  {Schreiber}, \citenamefont {Bloch},\ and\ \citenamefont
  {Schneider}}]{Bordia}%
  \BibitemOpen
  \bibfield  {author} {\bibinfo {author} {\bibfnamefont {Pranjal}\ \bibnamefont
  {Bordia}}, \bibinfo {author} {\bibfnamefont {Henrik~P.}\ \bibnamefont
  {L\"uschen}}, \bibinfo {author} {\bibfnamefont {Sean~S.}\ \bibnamefont
  {Hodgman}}, \bibinfo {author} {\bibfnamefont {Michael}\ \bibnamefont
  {Schreiber}}, \bibinfo {author} {\bibfnamefont {Immanuel}\ \bibnamefont
  {Bloch}}, \ and\ \bibinfo {author} {\bibfnamefont {Ulrich}\ \bibnamefont
  {Schneider}},\ }\bibfield  {title} {\enquote {\bibinfo {title} {Coupling
  identical one-dimensional many-body localized systems},}\ }\href {\doibase
  10.1103/PhysRevLett.116.140401} {\bibfield  {journal} {\bibinfo  {journal}
  {Phys. Rev. Lett.}\ }\textbf {\bibinfo {volume} {116}},\ \bibinfo {pages}
  {140401} (\bibinfo {year} {2016})}\BibitemShut {NoStop}%
\bibitem [{\citenamefont {Choi}\ \emph {et~al.}(2016)\citenamefont {Choi},
  \citenamefont {Hild}, \citenamefont {Zeiher}, \citenamefont {Schauss},
  \citenamefont {Rubio-Abadal}, \citenamefont {Yefsah}, \citenamefont
  {Khemani}, \citenamefont {Huse}, \citenamefont {Bloch},\ and\ \citenamefont
  {Gross}}]{Choi2016}%
  \BibitemOpen
  \bibfield  {author} {\bibinfo {author} {\bibfnamefont {J.}~\bibnamefont
  {Choi}}, \bibinfo {author} {\bibfnamefont {S.}~\bibnamefont {Hild}}, \bibinfo
  {author} {\bibfnamefont {J.}~\bibnamefont {Zeiher}}, \bibinfo {author}
  {\bibfnamefont {P.}~\bibnamefont {Schauss}}, \bibinfo {author} {\bibfnamefont
  {A.}~\bibnamefont {Rubio-Abadal}}, \bibinfo {author} {\bibfnamefont
  {T.}~\bibnamefont {Yefsah}}, \bibinfo {author} {\bibfnamefont
  {V.}~\bibnamefont {Khemani}}, \bibinfo {author} {\bibfnamefont {D.~A.}\
  \bibnamefont {Huse}}, \bibinfo {author} {\bibfnamefont {I.}~\bibnamefont
  {Bloch}}, \ and\ \bibinfo {author} {\bibfnamefont {C.}~\bibnamefont
  {Gross}},\ }\bibfield  {title} {\enquote {\bibinfo {title} {Exploring the
  many-body localization transition in two dimensions},}\ }\href {\doibase
  10.1126/science.aaf8834} {\bibfield  {journal} {\bibinfo  {journal}
  {Science}\ }\textbf {\bibinfo {volume} {352}},\ \bibinfo {pages} {1547--1552}
  (\bibinfo {year} {2016})}\BibitemShut {NoStop}%
\bibitem [{\citenamefont {Bardarson}\ \emph
  {et~al.}(2012{\natexlab{a}})\citenamefont {Bardarson}, \citenamefont
  {Pollmann},\ and\ \citenamefont {Moore}}]{Bardarson2012}%
  \BibitemOpen
  \bibfield  {author} {\bibinfo {author} {\bibfnamefont {Jens~H.}\ \bibnamefont
  {Bardarson}}, \bibinfo {author} {\bibfnamefont {Frank}\ \bibnamefont
  {Pollmann}}, \ and\ \bibinfo {author} {\bibfnamefont {Joel~E.}\ \bibnamefont
  {Moore}},\ }\bibfield  {title} {\enquote {\bibinfo {title} {Unbounded growth
  of entanglement in models of many-body localization},}\ }\href {\doibase
  10.1103/PhysRevLett.109.017202} {\bibfield  {journal} {\bibinfo  {journal}
  {Phys. Rev. Lett.}\ }\textbf {\bibinfo {volume} {109}},\ \bibinfo {pages}
  {017202} (\bibinfo {year} {2012}{\natexlab{a}})}\BibitemShut {NoStop}%
\bibitem [{\citenamefont {Serbyn}\ \emph {et~al.}(2013)\citenamefont {Serbyn},
  \citenamefont {Papi\ifmmode~\acute{c}\else \'{c}\fi{}},\ and\ \citenamefont
  {Abanin}}]{Serbyn}%
  \BibitemOpen
  \bibfield  {author} {\bibinfo {author} {\bibfnamefont {Maksym}\ \bibnamefont
  {Serbyn}}, \bibinfo {author} {\bibfnamefont {Z.}~\bibnamefont
  {Papi\ifmmode~\acute{c}\else \'{c}\fi{}}}, \ and\ \bibinfo {author}
  {\bibfnamefont {Dmitry~A.}\ \bibnamefont {Abanin}},\ }\bibfield  {title}
  {\enquote {\bibinfo {title} {Local conservation laws and the structure of the
  many-body localized states},}\ }\href {\doibase
  10.1103/PhysRevLett.111.127201} {\bibfield  {journal} {\bibinfo  {journal}
  {Phys. Rev. Lett.}\ }\textbf {\bibinfo {volume} {111}},\ \bibinfo {pages}
  {127201} (\bibinfo {year} {2013})}\BibitemShut {NoStop}%
\bibitem [{\citenamefont {Huse}\ \emph {et~al.}(2014)\citenamefont {Huse},
  \citenamefont {Nandkishore},\ and\ \citenamefont {Oganesyan}}]{HNO}%
  \BibitemOpen
  \bibfield  {author} {\bibinfo {author} {\bibfnamefont {David~A.}\
  \bibnamefont {Huse}}, \bibinfo {author} {\bibfnamefont {Rahul}\ \bibnamefont
  {Nandkishore}}, \ and\ \bibinfo {author} {\bibfnamefont {Vadim}\ \bibnamefont
  {Oganesyan}},\ }\bibfield  {title} {\enquote {\bibinfo {title} {Phenomenology
  of fully many-body-localized systems},}\ }\href {\doibase
  10.1103/PhysRevB.90.174202} {\bibfield  {journal} {\bibinfo  {journal} {Phys.
  Rev. B}\ }\textbf {\bibinfo {volume} {90}},\ \bibinfo {pages} {174202}
  (\bibinfo {year} {2014})}\BibitemShut {NoStop}%
\bibitem [{\citenamefont {Ros}\ \emph {et~al.}(2015)\citenamefont {Ros},
  \citenamefont {Müller},\ and\ \citenamefont {Scardicchio}}]{Scardicchio}%
  \BibitemOpen
  \bibfield  {author} {\bibinfo {author} {\bibfnamefont {V.}~\bibnamefont
  {Ros}}, \bibinfo {author} {\bibfnamefont {M.}~\bibnamefont {Müller}}, \ and\
  \bibinfo {author} {\bibfnamefont {A.}~\bibnamefont {Scardicchio}},\
  }\bibfield  {title} {\enquote {\bibinfo {title} {Integrals of motion in the
  many-body localized phase},}\ }\href {\doibase
  http://dx.doi.org/10.1016/j.nuclphysb.2014.12.014} {\bibfield  {journal}
  {\bibinfo  {journal} {Nuclear Physics B}\ }\textbf {\bibinfo {volume}
  {891}},\ \bibinfo {pages} {420 -- 465} (\bibinfo {year} {2015})}\BibitemShut
  {NoStop}%
\bibitem [{\citenamefont {Chandran}\ \emph {et~al.}()\citenamefont {Chandran},
  \citenamefont {Pal}, \citenamefont {Laumann},\ and\ \citenamefont
  {Scardicchio}}]{lstarbits}%
  \BibitemOpen
  \bibfield  {author} {\bibinfo {author} {\bibfnamefont {A.}~\bibnamefont
  {Chandran}}, \bibinfo {author} {\bibfnamefont {A.}~\bibnamefont {Pal}},
  \bibinfo {author} {\bibfnamefont {C.~R.}\ \bibnamefont {Laumann}}, \ and\
  \bibinfo {author} {\bibfnamefont {A.}~\bibnamefont {Scardicchio}},\
  }\href@noop {} {\bibinfo  {journal} {ArXiv e-prints: 1605.00655}\
  }\BibitemShut {NoStop}%
\bibitem [{\citenamefont {Geraedts}\ \emph
  {et~al.}(2017{\natexlab{a}})\citenamefont {Geraedts}, \citenamefont {Bhatt},\
  and\ \citenamefont {Nandkishore}}]{GBN}%
  \BibitemOpen
\bibfield  {journal} {  }\bibfield  {author} {\bibinfo {author} {\bibfnamefont
  {Scott~D.}\ \bibnamefont {Geraedts}}, \bibinfo {author} {\bibfnamefont
  {R.~N.}\ \bibnamefont {Bhatt}}, \ and\ \bibinfo {author} {\bibfnamefont
  {Rahul}\ \bibnamefont {Nandkishore}},\ }\bibfield  {title} {\enquote
  {\bibinfo {title} {Emergent local integrals of motion without a complete set
  of localized eigenstates},}\ }\href {\doibase 10.1103/PhysRevB.95.064204}
  {\bibfield  {journal} {\bibinfo  {journal} {Phys. Rev. B}\ }\textbf {\bibinfo
  {volume} {95}},\ \bibinfo {pages} {064204} (\bibinfo {year}
  {2017}{\natexlab{a}})}\BibitemShut {NoStop}%
\bibitem [{\citenamefont {{Parameswaran}}\ and\ \citenamefont
  {{Gopalakrishnan}}(2016)}]{NonFermiGlasses}%
  \BibitemOpen
  \bibfield  {author} {\bibinfo {author} {\bibfnamefont {S.~A.}\ \bibnamefont
  {{Parameswaran}}}\ and\ \bibinfo {author} {\bibfnamefont {S.}~\bibnamefont
  {{Gopalakrishnan}}},\ }\bibfield  {title} {\enquote {\bibinfo {title}
  {{Non-Fermi glasses: fractionalizing electrons at finite energy density}},}\
  }\href@noop {} {\bibfield  {journal} {\bibinfo  {journal} {ArXiv e-prints}\ }
  (\bibinfo {year} {2016})},\ \Eprint {http://arxiv.org/abs/1608.00981}
  {arXiv:1608.00981 [cond-mat.dis-nn]} \BibitemShut {NoStop}%
\bibitem [{\citenamefont {Khemani}\ \emph {et~al.}(2015)\citenamefont
  {Khemani}, \citenamefont {Nandkishore},\ and\ \citenamefont
  {Sondhi}}]{nonlocal}%
  \BibitemOpen
  \bibfield  {author} {\bibinfo {author} {\bibfnamefont {Vedika}\ \bibnamefont
  {Khemani}}, \bibinfo {author} {\bibfnamefont {Rahul}\ \bibnamefont
  {Nandkishore}}, \ and\ \bibinfo {author} {\bibfnamefont {S.~L.}\ \bibnamefont
  {Sondhi}},\ }\bibfield  {title} {\enquote {\bibinfo {title} {Nonlocal
  adiabatic response of a localized system to local manipulations},}\ }\href
  {http://dx.doi.org/10.1038/nphys3344} {\bibfield  {journal} {\bibinfo
  {journal} {Nat Phys}\ }\textbf {\bibinfo {volume} {11}},\ \bibinfo {pages}
  {560--565} (\bibinfo {year} {2015})}\BibitemShut {NoStop}%
\bibitem [{\citenamefont {Gopalakrishnan}\ \emph {et~al.}(2015)\citenamefont
  {Gopalakrishnan}, \citenamefont {M\"uller}, \citenamefont {Khemani},
  \citenamefont {Knap}, \citenamefont {Demler},\ and\ \citenamefont
  {Huse}}]{mblconductivity}%
  \BibitemOpen
  \bibfield  {author} {\bibinfo {author} {\bibfnamefont {Sarang}\ \bibnamefont
  {Gopalakrishnan}}, \bibinfo {author} {\bibfnamefont {Markus}\ \bibnamefont
  {M\"uller}}, \bibinfo {author} {\bibfnamefont {Vedika}\ \bibnamefont
  {Khemani}}, \bibinfo {author} {\bibfnamefont {Michael}\ \bibnamefont {Knap}},
  \bibinfo {author} {\bibfnamefont {Eugene}\ \bibnamefont {Demler}}, \ and\
  \bibinfo {author} {\bibfnamefont {David~A.}\ \bibnamefont {Huse}},\
  }\bibfield  {title} {\enquote {\bibinfo {title} {Low-frequency conductivity
  in many-body localized systems},}\ }\href {\doibase
  10.1103/PhysRevB.92.104202} {\bibfield  {journal} {\bibinfo  {journal} {Phys.
  Rev. B}\ }\textbf {\bibinfo {volume} {92}},\ \bibinfo {pages} {104202}
  (\bibinfo {year} {2015})}\BibitemShut {NoStop}%
\bibitem [{\citenamefont {Bardarson}\ \emph
  {et~al.}(2012{\natexlab{b}})\citenamefont {Bardarson}, \citenamefont
  {Pollmann},\ and\ \citenamefont {Moore}}]{Bardarson}%
  \BibitemOpen
  \bibfield  {author} {\bibinfo {author} {\bibfnamefont {Jens~H.}\ \bibnamefont
  {Bardarson}}, \bibinfo {author} {\bibfnamefont {Frank}\ \bibnamefont
  {Pollmann}}, \ and\ \bibinfo {author} {\bibfnamefont {Joel~E.}\ \bibnamefont
  {Moore}},\ }\bibfield  {title} {\enquote {\bibinfo {title} {Unbounded growth
  of entanglement in models of many-body localization},}\ }\href {\doibase
  10.1103/PhysRevLett.109.017202} {\bibfield  {journal} {\bibinfo  {journal}
  {Phys. Rev. Lett.}\ }\textbf {\bibinfo {volume} {109}},\ \bibinfo {pages}
  {017202} (\bibinfo {year} {2012}{\natexlab{b}})}\BibitemShut {NoStop}%
\bibitem [{\citenamefont {Geraedts}\ \emph {et~al.}(2016)\citenamefont
  {Geraedts}, \citenamefont {Nandkishore},\ and\ \citenamefont
  {Regnault}}]{Geraedts2016}%
  \BibitemOpen
  \bibfield  {author} {\bibinfo {author} {\bibfnamefont {Scott~D.}\
  \bibnamefont {Geraedts}}, \bibinfo {author} {\bibfnamefont {Rahul}\
  \bibnamefont {Nandkishore}}, \ and\ \bibinfo {author} {\bibfnamefont
  {Nicolas}\ \bibnamefont {Regnault}},\ }\bibfield  {title} {\enquote {\bibinfo
  {title} {Many-body localization and thermalization: Insights from the
  entanglement spectrum},}\ }\href {\doibase 10.1103/PhysRevB.93.174202}
  {\bibfield  {journal} {\bibinfo  {journal} {Phys. Rev. B}\ }\textbf {\bibinfo
  {volume} {93}},\ \bibinfo {pages} {174202} (\bibinfo {year}
  {2016})}\BibitemShut {NoStop}%
\bibitem [{\citenamefont {Khemani}\ \emph {et~al.}(2017)\citenamefont
  {Khemani}, \citenamefont {Lim}, \citenamefont {Sheng},\ and\ \citenamefont
  {Huse}}]{KhemaniPRX}%
  \BibitemOpen
  \bibfield  {author} {\bibinfo {author} {\bibfnamefont {Vedika}\ \bibnamefont
  {Khemani}}, \bibinfo {author} {\bibfnamefont {S.~P.}\ \bibnamefont {Lim}},
  \bibinfo {author} {\bibfnamefont {D.~N.}\ \bibnamefont {Sheng}}, \ and\
  \bibinfo {author} {\bibfnamefont {David~A.}\ \bibnamefont {Huse}},\
  }\bibfield  {title} {\enquote {\bibinfo {title} {Critical properties of the
  many-body localization transition},}\ }\href {\doibase
  10.1103/PhysRevX.7.021013} {\bibfield  {journal} {\bibinfo  {journal} {Phys.
  Rev. X}\ }\textbf {\bibinfo {volume} {7}},\ \bibinfo {pages} {021013}
  (\bibinfo {year} {2017})}\BibitemShut {NoStop}%
\bibitem [{\citenamefont {{Yang}}\ \emph {et~al.}(2017)\citenamefont {{Yang}},
  \citenamefont {{Hamma}}, \citenamefont {{Giampaolo}}, \citenamefont
  {{Mucciolo}},\ and\ \citenamefont {{Chamon}}}]{Chamon}%
  \BibitemOpen
  \bibfield  {author} {\bibinfo {author} {\bibfnamefont {Z.-C.}\ \bibnamefont
  {{Yang}}}, \bibinfo {author} {\bibfnamefont {A.}~\bibnamefont {{Hamma}}},
  \bibinfo {author} {\bibfnamefont {S.~M.}\ \bibnamefont {{Giampaolo}}},
  \bibinfo {author} {\bibfnamefont {E.~R.}\ \bibnamefont {{Mucciolo}}}, \ and\
  \bibinfo {author} {\bibfnamefont {C.}~\bibnamefont {{Chamon}}},\ }\bibfield
  {title} {\enquote {\bibinfo {title} {{Entanglement Complexity in Quantum
  Many-Body Dynamics, Thermalization and Localization}},}\ }\href@noop {}
  {\bibfield  {journal} {\bibinfo  {journal} {ArXiv e-prints}\ } (\bibinfo
  {year} {2017})},\ \Eprint {http://arxiv.org/abs/1703.03420} {arXiv:1703.03420
  [cond-mat.str-el]} \BibitemShut {NoStop}%
\bibitem [{\citenamefont {Geraedts}\ \emph
  {et~al.}(2017{\natexlab{b}})\citenamefont {Geraedts}, \citenamefont
  {Regnault},\ and\ \citenamefont {Nandkishore}}]{GRN}%
  \BibitemOpen
  \bibfield  {author} {\bibinfo {author} {\bibfnamefont {S~D}\ \bibnamefont
  {Geraedts}}, \bibinfo {author} {\bibfnamefont {N}~\bibnamefont {Regnault}}, \
  and\ \bibinfo {author} {\bibfnamefont {R~M}\ \bibnamefont {Nandkishore}},\
  }\bibfield  {title} {\enquote {\bibinfo {title} {Characterizing the many-body
  localization transition using the entanglement spectrum},}\ }\href
  {http://stacks.iop.org/1367-2630/19/i=11/a=113021} {\bibfield  {journal}
  {\bibinfo  {journal} {New Journal of Physics}\ }\textbf {\bibinfo {volume}
  {19}},\ \bibinfo {pages} {113021} (\bibinfo {year}
  {2017}{\natexlab{b}})}\BibitemShut {NoStop}%
\bibitem [{\citenamefont {Huse}\ \emph {et~al.}(2013)\citenamefont {Huse},
  \citenamefont {Nandkishore}, \citenamefont {Oganesyan}, \citenamefont {Pal},\
  and\ \citenamefont {Sondhi}}]{LPQO}%
  \BibitemOpen
  \bibfield  {author} {\bibinfo {author} {\bibfnamefont {David~A.}\
  \bibnamefont {Huse}}, \bibinfo {author} {\bibfnamefont {Rahul}\ \bibnamefont
  {Nandkishore}}, \bibinfo {author} {\bibfnamefont {Vadim}\ \bibnamefont
  {Oganesyan}}, \bibinfo {author} {\bibfnamefont {Arijeet}\ \bibnamefont
  {Pal}}, \ and\ \bibinfo {author} {\bibfnamefont {S.~L.}\ \bibnamefont
  {Sondhi}},\ }\bibfield  {title} {\enquote {\bibinfo {title}
  {Localization-protected quantum order},}\ }\href {\doibase
  10.1103/PhysRevB.88.014206} {\bibfield  {journal} {\bibinfo  {journal} {Phys.
  Rev. B}\ }\textbf {\bibinfo {volume} {88}},\ \bibinfo {pages} {014206}
  (\bibinfo {year} {2013})}\BibitemShut {NoStop}%
\bibitem [{\citenamefont {Pekker}\ \emph {et~al.}(2014)\citenamefont {Pekker},
  \citenamefont {Refael}, \citenamefont {Altman}, \citenamefont {Demler},\ and\
  \citenamefont {Oganesyan}}]{Pekkeretal2014}%
  \BibitemOpen
  \bibfield  {author} {\bibinfo {author} {\bibfnamefont {David}\ \bibnamefont
  {Pekker}}, \bibinfo {author} {\bibfnamefont {Gil}\ \bibnamefont {Refael}},
  \bibinfo {author} {\bibfnamefont {Ehud}\ \bibnamefont {Altman}}, \bibinfo
  {author} {\bibfnamefont {Eugene}\ \bibnamefont {Demler}}, \ and\ \bibinfo
  {author} {\bibfnamefont {Vadim}\ \bibnamefont {Oganesyan}},\ }\bibfield
  {title} {\enquote {\bibinfo {title} {Hilbert-glass transition: New
  universality of temperature-tuned many-body dynamical quantum criticality},}\
  }\href {\doibase 10.1103/PhysRevX.4.011052} {\bibfield  {journal} {\bibinfo
  {journal} {Phys. Rev. X}\ }\textbf {\bibinfo {volume} {4}},\ \bibinfo {pages}
  {011052} (\bibinfo {year} {2014})}\BibitemShut {NoStop}%
\bibitem [{\citenamefont {Vosk}\ and\ \citenamefont
  {Altman}(2014)}]{VoskAltman2014}%
  \BibitemOpen
  \bibfield  {author} {\bibinfo {author} {\bibfnamefont {Ronen}\ \bibnamefont
  {Vosk}}\ and\ \bibinfo {author} {\bibfnamefont {Ehud}\ \bibnamefont
  {Altman}},\ }\bibfield  {title} {\enquote {\bibinfo {title} {Dynamical
  quantum phase transitions in random spin chains},}\ }\href {\doibase
  10.1103/PhysRevLett.112.217204} {\bibfield  {journal} {\bibinfo  {journal}
  {Phys. Rev. Lett.}\ }\textbf {\bibinfo {volume} {112}},\ \bibinfo {pages}
  {217204} (\bibinfo {year} {2014})}\BibitemShut {NoStop}%
\bibitem [{\citenamefont {Nandkishore}\ and\ \citenamefont
  {Potter}(2014)}]{QHMBL}%
  \BibitemOpen
  \bibfield  {author} {\bibinfo {author} {\bibfnamefont {R.}~\bibnamefont
  {Nandkishore}}\ and\ \bibinfo {author} {\bibfnamefont {A.~C.}\ \bibnamefont
  {Potter}},\ }\bibfield  {title} {\enquote {\bibinfo {title} {Marginal
  anderson localization and many-body delocalization},}\ }\href {\doibase
  10.1103/PhysRevB.90.195115} {\bibfield  {journal} {\bibinfo  {journal} {Phys.
  Rev. B}\ }\textbf {\bibinfo {volume} {90}},\ \bibinfo {pages} {195115}
  (\bibinfo {year} {2014})}\BibitemShut {NoStop}%
\bibitem [{\citenamefont {Nandkishore}(2014)}]{2dcontinuum}%
  \BibitemOpen
  \bibfield  {author} {\bibinfo {author} {\bibfnamefont {Rahul}\ \bibnamefont
  {Nandkishore}},\ }\bibfield  {title} {\enquote {\bibinfo {title} {Many-body
  localization and delocalization in the two-dimensional continuum},}\ }\href
  {\doibase 10.1103/PhysRevB.90.184204} {\bibfield  {journal} {\bibinfo
  {journal} {Phys. Rev. B}\ }\textbf {\bibinfo {volume} {90}},\ \bibinfo
  {pages} {184204} (\bibinfo {year} {2014})}\BibitemShut {NoStop}%
\bibitem [{\citenamefont {{Gornyi}}\ \emph {et~al.}(2016)\citenamefont
  {{Gornyi}}, \citenamefont {{Mirlin}}, \citenamefont {{M{\"u}ller}},\ and\
  \citenamefont {{Polyakov}}}]{anycontinuum}%
  \BibitemOpen
  \bibfield  {author} {\bibinfo {author} {\bibfnamefont {I.~V.}\ \bibnamefont
  {{Gornyi}}}, \bibinfo {author} {\bibfnamefont {A.~D.}\ \bibnamefont
  {{Mirlin}}}, \bibinfo {author} {\bibfnamefont {M.}~\bibnamefont
  {{M{\"u}ller}}}, \ and\ \bibinfo {author} {\bibfnamefont {D.~G.}\
  \bibnamefont {{Polyakov}}},\ }\bibfield  {title} {\enquote {\bibinfo {title}
  {{Absence of many-body localization in a continuum}},}\ }\href@noop {}
  {\bibfield  {journal} {\bibinfo  {journal} {ArXiv e-prints}\ } (\bibinfo
  {year} {2016})},\ \Eprint {http://arxiv.org/abs/1611.05895} {arXiv:1611.05895
  [cond-mat.dis-nn]} \BibitemShut {NoStop}%
\bibitem [{\citenamefont {Nandkishore}(2015)}]{proximity}%
  \BibitemOpen
  \bibfield  {author} {\bibinfo {author} {\bibfnamefont {Rahul}\ \bibnamefont
  {Nandkishore}},\ }\bibfield  {title} {\enquote {\bibinfo {title} {Many-body
  localization proximity effect},}\ }\href {\doibase
  10.1103/PhysRevB.92.245141} {\bibfield  {journal} {\bibinfo  {journal} {Phys.
  Rev. B}\ }\textbf {\bibinfo {volume} {92}},\ \bibinfo {pages} {245141}
  (\bibinfo {year} {2015})}\BibitemShut {NoStop}%
\bibitem [{\citenamefont {Potter}\ and\ \citenamefont
  {Vasseur}(2016)}]{nonabelian}%
  \BibitemOpen
  \bibfield  {author} {\bibinfo {author} {\bibfnamefont {Andrew~C.}\
  \bibnamefont {Potter}}\ and\ \bibinfo {author} {\bibfnamefont {Romain}\
  \bibnamefont {Vasseur}},\ }\bibfield  {title} {\enquote {\bibinfo {title}
  {Symmetry constraints on many-body localization},}\ }\href {\doibase
  10.1103/PhysRevB.94.224206} {\bibfield  {journal} {\bibinfo  {journal} {Phys.
  Rev. B}\ }\textbf {\bibinfo {volume} {94}},\ \bibinfo {pages} {224206}
  (\bibinfo {year} {2016})}\BibitemShut {NoStop}%
\bibitem [{\citenamefont {De~Roeck}\ and\ \citenamefont
  {Huveneers}(2014)}]{DeRoeck2014}%
  \BibitemOpen
  \bibfield  {author} {\bibinfo {author} {\bibfnamefont {Wojciech}\
  \bibnamefont {De~Roeck}}\ and\ \bibinfo {author} {\bibfnamefont
  {Fran{\c{c}}ois}\ \bibnamefont {Huveneers}},\ }\bibfield  {title} {\enquote
  {\bibinfo {title} {Asymptotic quantum many-body localization from thermal
  disorder},}\ }\href {\doibase 10.1007/s00220-014-2116-8} {\bibfield
  {journal} {\bibinfo  {journal} {Communications in Mathematical Physics}\
  }\textbf {\bibinfo {volume} {332}},\ \bibinfo {pages} {1017--1082} (\bibinfo
  {year} {2014})}\BibitemShut {NoStop}%
\bibitem [{\citenamefont {De~Roeck}\ \emph {et~al.}(2016)\citenamefont
  {De~Roeck}, \citenamefont {Huveneers}, \citenamefont {M\"uller},\ and\
  \citenamefont {Schiulaz}}]{mblmobilityedges}%
  \BibitemOpen
  \bibfield  {author} {\bibinfo {author} {\bibfnamefont {Wojciech}\
  \bibnamefont {De~Roeck}}, \bibinfo {author} {\bibfnamefont {Francois}\
  \bibnamefont {Huveneers}}, \bibinfo {author} {\bibfnamefont {Markus}\
  \bibnamefont {M\"uller}}, \ and\ \bibinfo {author} {\bibfnamefont {Mauro}\
  \bibnamefont {Schiulaz}},\ }\bibfield  {title} {\enquote {\bibinfo {title}
  {Absence of many-body mobility edges},}\ }\href {\doibase
  10.1103/PhysRevB.93.014203} {\bibfield  {journal} {\bibinfo  {journal} {Phys.
  Rev. B}\ }\textbf {\bibinfo {volume} {93}},\ \bibinfo {pages} {014203}
  (\bibinfo {year} {2016})}\BibitemShut {NoStop}%
\bibitem [{\citenamefont {Nandkishore}\ and\ \citenamefont
  {Gopalakrishnan}()}]{mblbathgeneral}%
  \BibitemOpen
  \bibfield  {author} {\bibinfo {author} {\bibfnamefont {Rahul}\ \bibnamefont
  {Nandkishore}}\ and\ \bibinfo {author} {\bibfnamefont {Sarang}\ \bibnamefont
  {Gopalakrishnan}},\ }\bibfield  {title} {\enquote {\bibinfo {title} {Many
  body localized systems weakly coupled to baths},}\ }\href@noop {} {\bibinfo
  {journal} {Annalen der Physik,1521-3889 (2016)}\ }\BibitemShut {NoStop}%
\bibitem [{\citenamefont {{De Roeck}}\ and\ \citenamefont
  {{Huveneers}}(2016)}]{avalanches}%
  \BibitemOpen
\bibfield  {journal} {  }\bibfield  {author} {\bibinfo {author} {\bibfnamefont
  {W.}~\bibnamefont {{De Roeck}}}\ and\ \bibinfo {author} {\bibfnamefont
  {F.}~\bibnamefont {{Huveneers}}},\ }\bibfield  {title} {\enquote {\bibinfo
  {title} {{Stability and instability towards delocalization in MBL
  systems}},}\ }\href@noop {} {\bibfield  {journal} {\bibinfo  {journal} {ArXiv
  e-prints}\ } (\bibinfo {year} {2016})},\ \Eprint
  {http://arxiv.org/abs/1608.01815} {arXiv:1608.01815 [cond-mat.dis-nn]}
  \BibitemShut {NoStop}%
\bibitem [{\citenamefont {Parameswaran}\ and\ \citenamefont
  {Gopalakrishnan}(2017)}]{SILL}%
  \BibitemOpen
  \bibfield  {author} {\bibinfo {author} {\bibfnamefont {S.~A.}\ \bibnamefont
  {Parameswaran}}\ and\ \bibinfo {author} {\bibfnamefont {S.}~\bibnamefont
  {Gopalakrishnan}},\ }\bibfield  {title} {\enquote {\bibinfo {title}
  {Spin-catalyzed hopping conductivity in disordered strongly interacting
  quantum wires},}\ }\href {\doibase 10.1103/PhysRevB.95.024201} {\bibfield
  {journal} {\bibinfo  {journal} {Phys. Rev. B}\ }\textbf {\bibinfo {volume}
  {95}},\ \bibinfo {pages} {024201} (\bibinfo {year} {2017})}\BibitemShut
  {NoStop}%
\bibitem [{\citenamefont {Nandkishore}\ and\ \citenamefont
  {Sondhi}(2017)}]{LRMBL}%
  \BibitemOpen
  \bibfield  {author} {\bibinfo {author} {\bibfnamefont {Rahul~M.}\
  \bibnamefont {Nandkishore}}\ and\ \bibinfo {author} {\bibfnamefont {S.~L.}\
  \bibnamefont {Sondhi}},\ }\bibfield  {title} {\enquote {\bibinfo {title}
  {Many-body localization with long-range interactions},}\ }\href {\doibase
  10.1103/PhysRevX.7.041021} {\bibfield  {journal} {\bibinfo  {journal} {Phys.
  Rev. X}\ }\textbf {\bibinfo {volume} {7}},\ \bibinfo {pages} {041021}
  (\bibinfo {year} {2017})}\BibitemShut {NoStop}%
\bibitem [{\citenamefont {Nandkishore}\ \emph {et~al.}(2014)\citenamefont
  {Nandkishore}, \citenamefont {Gopalakrishnan},\ and\ \citenamefont
  {Huse}}]{NGH}%
  \BibitemOpen
  \bibfield  {author} {\bibinfo {author} {\bibfnamefont {Rahul}\ \bibnamefont
  {Nandkishore}}, \bibinfo {author} {\bibfnamefont {Sarang}\ \bibnamefont
  {Gopalakrishnan}}, \ and\ \bibinfo {author} {\bibfnamefont {David~A.}\
  \bibnamefont {Huse}},\ }\bibfield  {title} {\enquote {\bibinfo {title}
  {Spectral features of a many-body-localized system weakly coupled to a
  bath},}\ }\href {\doibase 10.1103/PhysRevB.90.064203} {\bibfield  {journal}
  {\bibinfo  {journal} {Phys. Rev. B}\ }\textbf {\bibinfo {volume} {90}},\
  \bibinfo {pages} {064203} (\bibinfo {year} {2014})}\BibitemShut {NoStop}%
\bibitem [{\citenamefont {Gopalakrishnan}\ and\ \citenamefont
  {Nandkishore}(2014)}]{gn}%
  \BibitemOpen
  \bibfield  {author} {\bibinfo {author} {\bibfnamefont {Sarang}\ \bibnamefont
  {Gopalakrishnan}}\ and\ \bibinfo {author} {\bibfnamefont {Rahul}\
  \bibnamefont {Nandkishore}},\ }\bibfield  {title} {\enquote {\bibinfo {title}
  {Mean-field theory of nearly many-body localized metals},}\ }\href {\doibase
  10.1103/PhysRevB.90.224203} {\bibfield  {journal} {\bibinfo  {journal} {Phys.
  Rev. B}\ }\textbf {\bibinfo {volume} {90}},\ \bibinfo {pages} {224203}
  (\bibinfo {year} {2014})}\BibitemShut {NoStop}%
\bibitem [{\citenamefont {Nandkishore}\ and\ \citenamefont
  {Gopalakrishnan}(2016)}]{NGADP}%
  \BibitemOpen
  \bibfield  {author} {\bibinfo {author} {\bibfnamefont {Rahul}\ \bibnamefont
  {Nandkishore}}\ and\ \bibinfo {author} {\bibfnamefont {Sarang}\ \bibnamefont
  {Gopalakrishnan}},\ }\bibfield  {title} {\enquote {\bibinfo {title} {Many
  body localized systems weakly coupled to baths},}\ }\href {\doibase
  10.1002/andp.201600181} {\bibfield  {journal} {\bibinfo  {journal} {Annalen
  der Physik}\ ,\ \bibinfo {pages} {1521--3889}} (\bibinfo {year}
  {2016})}\BibitemShut {NoStop}%
\bibitem [{\citenamefont {Banerjee}\ and\ \citenamefont
  {Altman}(2016)}]{BanerjeeAltman}%
  \BibitemOpen
  \bibfield  {author} {\bibinfo {author} {\bibfnamefont {Sumilan}\ \bibnamefont
  {Banerjee}}\ and\ \bibinfo {author} {\bibfnamefont {Ehud}\ \bibnamefont
  {Altman}},\ }\bibfield  {title} {\enquote {\bibinfo {title} {Variable-range
  hopping through marginally localized phonons},}\ }\href {\doibase
  10.1103/PhysRevLett.116.116601} {\bibfield  {journal} {\bibinfo  {journal}
  {Phys. Rev. Lett.}\ }\textbf {\bibinfo {volume} {116}},\ \bibinfo {pages}
  {116601} (\bibinfo {year} {2016})}\BibitemShut {NoStop}%
\bibitem [{\citenamefont {Fischer}\ \emph {et~al.}(2016)\citenamefont
  {Fischer}, \citenamefont {Maksymenko},\ and\ \citenamefont
  {Altman}}]{Altmandeph}%
  \BibitemOpen
  \bibfield  {author} {\bibinfo {author} {\bibfnamefont {Mark~H}\ \bibnamefont
  {Fischer}}, \bibinfo {author} {\bibfnamefont {Mykola}\ \bibnamefont
  {Maksymenko}}, \ and\ \bibinfo {author} {\bibfnamefont {Ehud}\ \bibnamefont
  {Altman}},\ }\bibfield  {title} {\enquote {\bibinfo {title} {Dynamics of a
  many-body-localized system coupled to a bath},}\ }\href {\doibase
  10.1103/PhysRevLett.116.160401} {\bibfield  {journal} {\bibinfo  {journal}
  {Phys. Rev. Lett.}\ }\textbf {\bibinfo {volume} {116}},\ \bibinfo {pages}
  {160401} (\bibinfo {year} {2016})}\BibitemShut {NoStop}%
\bibitem [{\citenamefont {Levi}\ \emph {et~al.}(2016)\citenamefont {Levi},
  \citenamefont {Heyl}, \citenamefont {Lesanovsky},\ and\ \citenamefont
  {Garrahan}}]{Levi}%
  \BibitemOpen
  \bibfield  {author} {\bibinfo {author} {\bibfnamefont {Emanuele}\
  \bibnamefont {Levi}}, \bibinfo {author} {\bibfnamefont {Markus}\ \bibnamefont
  {Heyl}}, \bibinfo {author} {\bibfnamefont {Igor}\ \bibnamefont {Lesanovsky}},
  \ and\ \bibinfo {author} {\bibfnamefont {Juan~P.}\ \bibnamefont {Garrahan}},\
  }\bibfield  {title} {\enquote {\bibinfo {title} {Robustness of many-body
  localization in the presence of dissipation},}\ }\href {\doibase
  10.1103/PhysRevLett.116.237203} {\bibfield  {journal} {\bibinfo  {journal}
  {Phys. Rev. Lett.}\ }\textbf {\bibinfo {volume} {116}},\ \bibinfo {pages}
  {237203} (\bibinfo {year} {2016})}\BibitemShut {NoStop}%
\bibitem [{\citenamefont {Medvedyeva}\ \emph {et~al.}(2016)\citenamefont
  {Medvedyeva}, \citenamefont {Prosen},\ and\ \citenamefont
  {Znidaric}}]{Medv16}%
  \BibitemOpen
  \bibfield  {author} {\bibinfo {author} {\bibfnamefont {M.}~\bibnamefont
  {Medvedyeva}}, \bibinfo {author} {\bibfnamefont {T.}~\bibnamefont {Prosen}},
  \ and\ \bibinfo {author} {\bibfnamefont {M.}~\bibnamefont {Znidaric}},\
  }\bibfield  {title} {\enquote {\bibinfo {title} {Influence of dephasing on
  many-body localization},}\ }\href {\doibase 10.1103/PhysRevB.93.094205}
  {\bibfield  {journal} {\bibinfo  {journal} {Phys. Rev. B}\ }\textbf {\bibinfo
  {volume} {93}},\ \bibinfo {pages} {094205} (\bibinfo {year}
  {2016})}\BibitemShut {NoStop}%
\bibitem [{\citenamefont {Luitz}\ \emph {et~al.}(2017)\citenamefont {Luitz},
  \citenamefont {Huveneers},\ and\ \citenamefont {De~Roeck}}]{deroeck17}%
  \BibitemOpen
  \bibfield  {author} {\bibinfo {author} {\bibfnamefont {D.~J.}\ \bibnamefont
  {Luitz}}, \bibinfo {author} {\bibfnamefont {F.}~\bibnamefont {Huveneers}}, \
  and\ \bibinfo {author} {\bibfnamefont {W.}~\bibnamefont {De~Roeck}},\
  }\bibfield  {title} {\enquote {\bibinfo {title} {How a small quantum bath can
  thermalize long localized chains},}\ }\href {\doibase
  10.1103/PhysRevLett.119.150602} {\bibfield  {journal} {\bibinfo  {journal}
  {Phys. Rev. Lett.}\ }\textbf {\bibinfo {volume} {119}},\ \bibinfo {pages}
  {150602} (\bibinfo {year} {2017})}\BibitemShut {NoStop}%
\bibitem [{\citenamefont {Ponte}\ \emph {et~al.}(2017)\citenamefont {Ponte},
  \citenamefont {Laumann}, \citenamefont {Huse},\ and\ \citenamefont
  {Chandran}}]{Chandran17}%
  \BibitemOpen
  \bibfield  {author} {\bibinfo {author} {\bibfnamefont {P.}~\bibnamefont
  {Ponte}}, \bibinfo {author} {\bibfnamefont {C.}~\bibnamefont {Laumann}},
  \bibinfo {author} {\bibfnamefont {D.}~\bibnamefont {Huse}}, \ and\ \bibinfo
  {author} {\bibfnamefont {A.}~\bibnamefont {Chandran}},\ }\bibfield  {title}
  {\enquote {\bibinfo {title} {Thermal inclusions: how one spin can destroy a
  many-body localized phase},}\ }\href@noop {} {\bibfield  {journal} {\bibinfo
  {journal} {Philosophical Transactions A}\ } (\bibinfo {year}
  {2017})}\BibitemShut {NoStop}%
\bibitem [{\citenamefont {L\"uschen}\ \emph {et~al.}(2017)\citenamefont
  {L\"uschen}, \citenamefont {Bordia}, \citenamefont {Hodgman}, \citenamefont
  {Schreiber}, \citenamefont {Sarkar}, \citenamefont {Daley}, \citenamefont
  {Fischer}, \citenamefont {Altman}, \citenamefont {Bloch},\ and\ \citenamefont
  {Schneider}}]{Bloch17}%
  \BibitemOpen
  \bibfield  {author} {\bibinfo {author} {\bibfnamefont {H.~P.}\ \bibnamefont
  {L\"uschen}}, \bibinfo {author} {\bibfnamefont {P.}~\bibnamefont {Bordia}},
  \bibinfo {author} {\bibfnamefont {S.~S.}\ \bibnamefont {Hodgman}}, \bibinfo
  {author} {\bibfnamefont {M.}~\bibnamefont {Schreiber}}, \bibinfo {author}
  {\bibfnamefont {S.}~\bibnamefont {Sarkar}}, \bibinfo {author} {\bibfnamefont
  {A.~J.}\ \bibnamefont {Daley}}, \bibinfo {author} {\bibfnamefont {M.~H.}\
  \bibnamefont {Fischer}}, \bibinfo {author} {\bibfnamefont {E.}~\bibnamefont
  {Altman}}, \bibinfo {author} {\bibfnamefont {I.}~\bibnamefont {Bloch}}, \
  and\ \bibinfo {author} {\bibfnamefont {U.}~\bibnamefont {Schneider}},\
  }\bibfield  {title} {\enquote {\bibinfo {title} {Signatures of many-body
  localization in a controlled open quantum system},}\ }\href {\doibase
  10.1103/PhysRevX.7.011034} {\bibfield  {journal} {\bibinfo  {journal} {Phys.
  Rev. X}\ }\textbf {\bibinfo {volume} {7}},\ \bibinfo {pages} {011034}
  (\bibinfo {year} {2017})}\BibitemShut {NoStop}%
\bibitem [{\citenamefont {Imbrie}\ \emph {et~al.}(2016)\citenamefont {Imbrie},
  \citenamefont {Ros},\ and\ \citenamefont {Scardicchio}}]{Imbriereview}%
  \BibitemOpen
  \bibfield  {author} {\bibinfo {author} {\bibfnamefont {J.~Z}\ \bibnamefont
  {Imbrie}}, \bibinfo {author} {\bibfnamefont {V.}~\bibnamefont {Ros}}, \ and\
  \bibinfo {author} {\bibfnamefont {A.}~\bibnamefont {Scardicchio}},\
  }\href@noop {} {\bibfield  {journal} {\bibinfo  {journal} {arXiv:cond-mat
  1609.08076}\ } (\bibinfo {year} {2016})}\BibitemShut {NoStop}%
\bibitem [{\citenamefont {Breuer}\ and\ \citenamefont
  {Petruccione}(2007)}]{Breuerbook}%
  \BibitemOpen
  \bibfield  {author} {\bibinfo {author} {\bibfnamefont {H-P}\ \bibnamefont
  {Breuer}}\ and\ \bibinfo {author} {\bibfnamefont {F.}~\bibnamefont
  {Petruccione}},\ }\href@noop {} {\emph {\bibinfo {title} {The Theory of Open
  Quantum Systems}}}\ (\bibinfo  {publisher} {Oxford University Press},\
  \bibinfo {year} {2007})\BibitemShut {NoStop}%
\bibitem [{\citenamefont {Marcos}\ \emph {et~al.}()\citenamefont {Marcos},
  \citenamefont {Tomadin}, \citenamefont {Diehl},\ and\ \citenamefont
  {Rabl}}]{Marcos12}%
  \BibitemOpen
  \bibfield  {author} {\bibinfo {author} {\bibfnamefont {D.}~\bibnamefont
  {Marcos}}, \bibinfo {author} {\bibfnamefont {A.}~\bibnamefont {Tomadin}},
  \bibinfo {author} {\bibfnamefont {S.}~\bibnamefont {Diehl}}, \ and\ \bibinfo
  {author} {\bibfnamefont {P.}~\bibnamefont {Rabl}},\ }\bibfield  {title}
  {\enquote {\bibinfo {title} {Photon condensation in circuit qed by engineered
  dissipation},}\ }\href@noop {} {\bibinfo  {journal} {New J. Phys. 14, 055005
  (2012)}\ }\BibitemShut {NoStop}%
\bibitem [{\citenamefont {Agarwal}\ \emph {et~al.}(1997)\citenamefont
  {Agarwal}, \citenamefont {Puri},\ and\ \citenamefont {Singh}}]{Agarw}%
  \BibitemOpen
\bibfield  {journal} {  }\bibfield  {author} {\bibinfo {author} {\bibfnamefont
  {G.~S.}\ \bibnamefont {Agarwal}}, \bibinfo {author} {\bibfnamefont {R.~R.}\
  \bibnamefont {Puri}}, \ and\ \bibinfo {author} {\bibfnamefont {R.~P.}\
  \bibnamefont {Singh}},\ }\bibfield  {title} {\enquote {\bibinfo {title}
  {Atomic schr\"odinger cat states},}\ }\href {\doibase
  10.1103/PhysRevA.56.2249} {\bibfield  {journal} {\bibinfo  {journal} {Phys.
  Rev. A}\ }\textbf {\bibinfo {volume} {56}},\ \bibinfo {pages} {2249--2254}
  (\bibinfo {year} {1997})}\BibitemShut {NoStop}%
\bibitem [{\citenamefont {Houck}\ \emph {et~al.}(2012)\citenamefont {Houck},
  \citenamefont {Tureci},\ and\ \citenamefont {Koch}}]{houck12}%
  \BibitemOpen
  \bibfield  {author} {\bibinfo {author} {\bibfnamefont {A.}~\bibnamefont
  {Houck}}, \bibinfo {author} {\bibfnamefont {H.~E.}\ \bibnamefont {Tureci}}, \
  and\ \bibinfo {author} {\bibfnamefont {J.}~\bibnamefont {Koch}},\ }\bibfield
  {title} {\enquote {\bibinfo {title} {On-chip quantum simulation with
  superconducting circuits},}\ }\href@noop {} {\bibfield  {journal} {\bibinfo
  {journal} {Nature Physics}\ }\textbf {\bibinfo {volume} {8}},\ \bibinfo
  {pages} {292 -- 299} (\bibinfo {year} {2012})}\BibitemShut {NoStop}%
\bibitem [{\citenamefont {Underwood}\ \emph {et~al.}(2012)\citenamefont
  {Underwood}, \citenamefont {Shanks}, \citenamefont {Koch},\ and\
  \citenamefont {Houck}}]{Underwood12}%
  \BibitemOpen
  \bibfield  {author} {\bibinfo {author} {\bibfnamefont {D.~L.}\ \bibnamefont
  {Underwood}}, \bibinfo {author} {\bibfnamefont {W.~E.}\ \bibnamefont
  {Shanks}}, \bibinfo {author} {\bibfnamefont {J.}~\bibnamefont {Koch}}, \ and\
  \bibinfo {author} {\bibfnamefont {A.~A.}\ \bibnamefont {Houck}},\ }\bibfield
  {title} {\enquote {\bibinfo {title} {Low-disorder microwave cavity lattices
  for quantum simulation with photons},}\ }\href {\doibase
  10.1103/PhysRevA.86.023837} {\bibfield  {journal} {\bibinfo  {journal} {Phys.
  Rev. A}\ }\textbf {\bibinfo {volume} {86}},\ \bibinfo {pages} {023837}
  (\bibinfo {year} {2012})}\BibitemShut {NoStop}%
\bibitem [{\citenamefont {Fitzpatrick}\ \emph {et~al.}(2017)\citenamefont
  {Fitzpatrick}, \citenamefont {Sundaresan}, \citenamefont {Li}, \citenamefont
  {Koch},\ and\ \citenamefont {Houck}}]{Kirk17}%
  \BibitemOpen
  \bibfield  {author} {\bibinfo {author} {\bibfnamefont {M.}~\bibnamefont
  {Fitzpatrick}}, \bibinfo {author} {\bibfnamefont {N.~M.}\ \bibnamefont
  {Sundaresan}}, \bibinfo {author} {\bibfnamefont {Andy C.~Y.}\ \bibnamefont
  {Li}}, \bibinfo {author} {\bibfnamefont {J.}~\bibnamefont {Koch}}, \ and\
  \bibinfo {author} {\bibfnamefont {A.~A.}\ \bibnamefont {Houck}},\ }\bibfield
  {title} {\enquote {\bibinfo {title} {Observation of a dissipative phase
  transition in a one-dimensional circuit qed lattice},}\ }\href {\doibase
  10.1103/PhysRevX.7.011016} {\bibfield  {journal} {\bibinfo  {journal} {Phys.
  Rev. X}\ }\textbf {\bibinfo {volume} {7}},\ \bibinfo {pages} {011016}
  (\bibinfo {year} {2017})}\BibitemShut {NoStop}%
\bibitem [{\citenamefont {Hartmann}\ \emph {et~al.}()\citenamefont {Hartmann},
  \citenamefont {Brandao},\ and\ \citenamefont {Plenio}}]{Hartmann}%
  \BibitemOpen
  \bibfield  {author} {\bibinfo {author} {\bibfnamefont {M.}~\bibnamefont
  {Hartmann}}, \bibinfo {author} {\bibfnamefont {F.}~\bibnamefont {Brandao}}, \
  and\ \bibinfo {author} {\bibfnamefont {M.}~\bibnamefont {Plenio}},\
  }\bibfield  {title} {\enquote {\bibinfo {title} {Quantum many-body phenomena
  in coupled cavity arrays},}\ }\href@noop {} {\bibinfo  {journal} {Laser
  Photonics Rev. 2, 527}\ }\BibitemShut {NoStop}%
\bibitem [{\citenamefont {Tomadin}\ and\ \citenamefont {Fazio}()}]{Tomadin}%
  \BibitemOpen
\bibfield  {journal} {  }\bibfield  {author} {\bibinfo {author} {\bibfnamefont
  {A.}~\bibnamefont {Tomadin}}\ and\ \bibinfo {author} {\bibfnamefont
  {R.}~\bibnamefont {Fazio}},\ }\bibfield  {title} {\enquote {\bibinfo {title}
  {Many-body phenomena in qed-cavity arrays},}\ }\href@noop {} {\bibinfo
  {journal} {J. Opt. Soc. Am. B, 27, A130 (2010)}\ }\BibitemShut {NoStop}%
\bibitem [{\citenamefont {Chang}\ \emph {et~al.}(2007)\citenamefont {Chang},
  \citenamefont {Sorensen}, \citenamefont {Demler},\ and\ \citenamefont
  {Lukin}}]{Chang}%
  \BibitemOpen
\bibfield  {journal} {  }\bibfield  {author} {\bibinfo {author} {\bibfnamefont
  {D.~E}\ \bibnamefont {Chang}}, \bibinfo {author} {\bibfnamefont
  {A.}~\bibnamefont {Sorensen}}, \bibinfo {author} {\bibfnamefont
  {E.}~\bibnamefont {Demler}}, \ and\ \bibinfo {author} {\bibfnamefont
  {M.}~\bibnamefont {Lukin}},\ }\bibfield  {title} {\enquote {\bibinfo {title}
  {A single-photon transistor using nanoscale surface plasmons},}\ }\href@noop
  {} {\bibfield  {journal} {\bibinfo  {journal} {Nature Physics, 3, 807-812}\ }
  (\bibinfo {year} {2007})}\BibitemShut {NoStop}%
\bibitem [{\citenamefont {Zeiher}\ and\ \citenamefont {\emph{et
  al.}}(2016)}]{spinrdyd}%
  \BibitemOpen
  \bibfield  {author} {\bibinfo {author} {\bibfnamefont {J.}~\bibnamefont
  {Zeiher}}\ and\ \bibinfo {author} {\bibnamefont {\emph{et al.}}},\ }\bibfield
   {title} {\enquote {\bibinfo {title} {Many-body interferometry of a
  rydberg-dressed spin lattice},}\ }\href@noop {} {\bibfield  {journal}
  {\bibinfo  {journal} {Nature Physics, 12, 1095-1099}\ } (\bibinfo {year}
  {2016})}\BibitemShut {NoStop}%
\bibitem [{\citenamefont {Pekker}\ \emph {et~al.}(2017)\citenamefont {Pekker},
  \citenamefont {Clark}, \citenamefont {Oganesyan},\ and\ \citenamefont
  {Refael}}]{Pekker17}%
  \BibitemOpen
  \bibfield  {author} {\bibinfo {author} {\bibfnamefont {David}\ \bibnamefont
  {Pekker}}, \bibinfo {author} {\bibfnamefont {Bryan~K.}\ \bibnamefont
  {Clark}}, \bibinfo {author} {\bibfnamefont {Vadim}\ \bibnamefont
  {Oganesyan}}, \ and\ \bibinfo {author} {\bibfnamefont {Gil}\ \bibnamefont
  {Refael}},\ }\bibfield  {title} {\enquote {\bibinfo {title} {Fixed points of
  wegner-wilson flows and many-body localization},}\ }\href {\doibase
  10.1103/PhysRevLett.119.075701} {\bibfield  {journal} {\bibinfo  {journal}
  {Phys. Rev. Lett.}\ }\textbf {\bibinfo {volume} {119}},\ \bibinfo {pages}
  {075701} (\bibinfo {year} {2017})}\BibitemShut {NoStop}%
\bibitem [{\citenamefont {Thomson}\ and\ \citenamefont
  {Schiro}(2017)}]{Thomson17}%
  \BibitemOpen
  \bibfield  {author} {\bibinfo {author} {\bibfnamefont {S.}~\bibnamefont
  {Thomson}}\ and\ \bibinfo {author} {\bibfnamefont {M.}~\bibnamefont
  {Schiro}},\ }\href@noop {} {\bibfield  {journal} {\bibinfo  {journal}
  {arXiv:cond-mat}\ ,\ \bibinfo {pages} {1707.06981}} (\bibinfo {year}
  {2017})}\BibitemShut {NoStop}%
\bibitem [{\citenamefont {Dasgupta}\ and\ \citenamefont {Ma}(1980)}]{das}%
  \BibitemOpen
  \bibfield  {author} {\bibinfo {author} {\bibfnamefont {Chandan}\ \bibnamefont
  {Dasgupta}}\ and\ \bibinfo {author} {\bibfnamefont {Shang-keng}\ \bibnamefont
  {Ma}},\ }\bibfield  {title} {\enquote {\bibinfo {title} {Low-temperature
  properties of the random heisenberg antiferromagnetic chain},}\ }\href
  {\doibase 10.1103/PhysRevB.22.1305} {\bibfield  {journal} {\bibinfo
  {journal} {Phys. Rev. B}\ }\textbf {\bibinfo {volume} {22}},\ \bibinfo
  {pages} {1305--1319} (\bibinfo {year} {1980})}\BibitemShut {NoStop}%
\bibitem [{\citenamefont {Fisher}(1994)}]{fis}%
  \BibitemOpen
  \bibfield  {author} {\bibinfo {author} {\bibfnamefont {Daniel~S.}\
  \bibnamefont {Fisher}},\ }\bibfield  {title} {\enquote {\bibinfo {title}
  {Random antiferromagnetic quantum spin chains},}\ }\href {\doibase
  10.1103/PhysRevB.50.3799} {\bibfield  {journal} {\bibinfo  {journal} {Phys.
  Rev. B}\ }\textbf {\bibinfo {volume} {50}},\ \bibinfo {pages} {3799--3821}
  (\bibinfo {year} {1994})}\BibitemShut {NoStop}%
\bibitem [{\citenamefont {Vosk}\ and\ \citenamefont {Altman}(2013)}]{vosk}%
  \BibitemOpen
  \bibfield  {author} {\bibinfo {author} {\bibfnamefont {Ronen}\ \bibnamefont
  {Vosk}}\ and\ \bibinfo {author} {\bibfnamefont {Ehud}\ \bibnamefont
  {Altman}},\ }\bibfield  {title} {\enquote {\bibinfo {title} {Many-body
  localization in one dimension as a dynamical renormalization group fixed
  point},}\ }\href {\doibase 10.1103/PhysRevLett.110.067204} {\bibfield
  {journal} {\bibinfo  {journal} {Phys. Rev. Lett.}\ }\textbf {\bibinfo
  {volume} {110}},\ \bibinfo {pages} {067204} (\bibinfo {year}
  {2013})}\BibitemShut {NoStop}%
\end{thebibliography}%

\newpage
\begin{widetext}

\end{widetext}

 \end{document}